\documentclass[prb,aps,twocolumn,showpacs,floatfix]{revtex4-1}
\usepackage{amsfonts}
\usepackage{amssymb}
\usepackage{hyperref}
\usepackage{graphicx}
\usepackage{dcolumn}
\usepackage{bm,amsmath,verbatim}
\usepackage{mathrsfs}
\usepackage{color}

\hypersetup{colorlinks,
linkcolor=blue,          citecolor=blue,        filecolor=blue,      urlcolor=blue           }

\def\be{\begin{equation}}
\def\ee{\end{equation}}
\def\bea{\begin{eqnarray}}
\def\eea{\end{eqnarray}}
\def\bse{\begin{subequations}}
\def\ese{\end{subequations}}

\def\be{\begin{eqnarray}}
\def\ee{\end{eqnarray}}

\begin{document}

\title{Topological Gapless Matters in Three-dimensional Ultracold Atomic Gases}
\author{Yong Xu}
\email{yongxuphy@tsinghua.edu.cn}
\affiliation{Center for Quantum Information, IIIS, Tsinghua University, Beijing 100084, PR China}

\begin{abstract}
Three-dimensional topological gapless matters with gapless degeneracies protected by a topological invariant defined over a
closed manifold in momentum space have attracted considerable interest in various fields ranging from condensed matter materials to ultracold atomic gases. As a highly controllable and disorder free system,
ultracold atomic gases provide a versatile platform to simulate topological gapless matters. Here, the current
progress in studies of topological gapless phenomena in three-dimensional cold atom systems is summarized in the review.
It is mainly focused on Weyl points, structured (type-II) Weyl points, Dirac points, nodal rings and Weyl exceptional rings in cold atoms. Since interactions in cold atoms can be controlled via Feshbach resonances, the progress in both superfluids
for attractive interactions and non-interacting cold atom gases is reviewed.
\end{abstract}

\maketitle

\tableofcontents.


\section{Introduction}
In 1928, P. Dirac proposed the Dirac equation to describe particles with Lorentz invariance~\cite{Dirac}.
Shortly afterward, H. Weyl found that, for massless particles, the equation can be written as
two separate equations corresponding to the left or right chirality or handedness~\cite{Weyl}. The
fermions described by these two equations are called Weyl fermions. Initially, neutrinos
were thought to be Weyl fermions, but they were found
to have masses and thus not Weyl fermions. So far, there are no other elementary particles
in particle physics that are Weyl fermions.

In condensed matter systems including cold atoms,
particles do not necessarily endorse the law dictated by the Dirac equation
due to their low energy compared with their static masses. However, the existence of band structures
due to crystalline symmetries or atom laser interactions can dress the particles, leading
to their dynamics governed by the Weyl equation. This indeed occurs when a band structure exhibits
a doubly degenerate point with linear dispersions along all three dimensions in the vicinity
of the point. Around such a point (called Weyl node or Weyl point), the dynamics is dictated by the Weyl Hamiltonian.
One celebrated example in two dimensions (2D) is graphene where a doubly degenerate point appears~\cite{Geim2009RMP,DasSarma2011RMP}. In fact, as early as in 1937, such a degeneracy
with linear dispersions was predicted in electric band structures in a three-dimensional (3D) solid-state material~\cite{Herring}.

Half a century later, Volovik predicted that the Weyl point can appear in
quasiparticle spectra of the $^{3}$He A superfluid phase~\cite{Volovik1987,volovik}. It was not until in 2011 that
Wan and coworkers discovered the topological consequence of materials with Weyl points:
Fermi arcs that consist of surface states connecting two Weyl points with opposite chirality~\cite{Wan2011prb}.
The appearance of Fermi arcs is due to the topological property of Weyl points:
Weyl points can be viewed as the monopole of Berry curvatures in momentum space so that
the first Chern number defined as the integral of Berry curvatures over a closed surface in momentum
space enclosing the point is quantized. From this perspective, the closed surface can be regarded as a Chern
insulator, leading to chiral surface states in an open geometry; these surface states
give rise to a Fermi arc connecting two Weyl points. So far, studies of Weyl points have
seen a rapid advance in various fields, such as solid-state materials~\cite{Wan2011prb,Yuanming2011PRB,Burkov2011PRL,ZhongFang2011prl,Bernevig2012PRL,Shengyuan2014PRL,Weng2015PRX,Ueda2016,FuLiang2016NP},
cold atoms~\cite{Gong2011prl,JianHua2012PRA,Anderson2012PRL,Xu2014PRL,YongReveiw,Liubo2015PRL,sarma2015PRB,Xiaopeng2015NC,Tena2015RPL,
Xu2015PRL,KTLaw2016,XJLiu2016PRA,Yong2016PRATypeII,Ray2015NC,Lepori2016PRB,Lepori2016PRA,DanWei2017PRA,Lijun2018PRB,Xiongjun2018PRA,Yanbin2018PRB,Chuanwei2018arXiv,YongHu2018},
optical systems~\cite{LingLu2013NP}, acoustic systems~\cite{Meng2015} and mechanical systems~\cite{Rocklin2016RPL}.
Remarkably, Weyl points have recently been
experimentally observed in solid-state materials~\cite{Lv2015,Xu2015} and optical crystals~\cite{Lu2015,Wenjie2016nc,Biao2018science}.

In contrast to particle physics where the Lorentz invariance is required,
in condensed matter systems, this symmetry is not necessarily respected. As a result, new fermions
can appear. In 2014, Xu and coworkers found a new topological superfluid termed
gapless topological Fulde-Ferrell superfluids; these superfluids possess gapless structures
besides a touching point~\cite{Xu2014PRL}. Shortly afterward, Xu and coworkers proposed an effective model
to characterize it and named the new Weyl point structured Weyl points~\cite{Xu2015PRL}.
Such Weyl points were also found in solid-state materials and named type-II Weyl points~\cite{Soluyanov2015Nature,Bergholtz2015PRL}.
For a traditional Weyl point (also called type-I Weyl point), the Fermi surface is a point with zero density of
states when it lies at the point. However, for a structured (type-II) Weyl point,
the Fermi surface incorporates two open or closed surfaces touching at the point, leading to
finite density of states. Because of the sharp change of the topology of the Fermi surface,
the transition from the traditional (type-I) Weyl point to the structured (type-II) Weyl point
corresponds to the Lifshitz transition. Besides, a number of other different properties of structured (type-II)
Weyl points were anticipated, such as anisotropic chiral anomaly~\cite{Soluyanov2015Nature,Yang2016,Udagawa2016},
unusual magneto-response~\cite{Yang2016,Udagawa2016},
novel quantum oscillation due to the momentum space Klein tunneling~\cite{Beenakker2016PRL}, and
novel anomalous Hall effects~\cite{Xu2015PRL,Zyuzin2018JETP}. Such new Weyl points have recently been
experimentally observed in solid-state materials~\cite{Shuyun2016,Huang2016,Liang2016arxiv,Tamai2016PRX,Hasan2017ScienceAdv,Jiang2017nc,Liang2016,NXu2016,Bruno2016PRB} and optical crystals~\cite{Rechtsman2017NP,Biao2017nc}.

In addition to the point, 3D band structures can exhibit nodal lines (or rings) with
degeneracies forming a line (or a ring)~\cite{Burkov2011PRB}(see also review~\cite{ChenFang2016CPB} and
references therein). For instance, a nodal line (or ring) can be
developed from Dirac points in graphene by generalizing them to 3D. Such a gapless structure is
protected by the winding number for a system with chiral symmetry~\cite{Burkov2011PRB} or the quantized Berry phase
for a system with $PT$ symmetry~\cite{Kim2015PRL}.
Moreover, other nodal lines, such as a nodal line acting as a vortex ring with a maximal topological anomalous
Hall effect~\cite{Lim2017PRL}, nodal-link and nodal-knot gapless structures~\cite{Wei2017PRB,Zhongbo2017PRB,Yee2017PRB,Ren2017PRB}, are proposed
in band structures of different models.

Topological gapless matters can not only appear in closed systems described by a Hermitian Hamiltonian,
but can also emerge in open systems with particle gain and loss described by a non-Hermitian Hamiltonian.
Recently, the technology advance~\cite{Kip2010,Yang2014,Zhang2014Science,Khajavikhan2014Science,Zeuner2015PRL,Moljacic2015,Gao2015,Harris2016nature,LuoleArxiv,Weimann,Weijian2017nature,
Xue2017NP,Hengyun2018science,Rechtsman2018} has instigated intensive studies of topological phases in non-Hermitian systems ranging from 1D to 3D systems~\cite{Rudner1,Esaki,Bardyn,Malzard,Lee3,Zeng,Leykam,Xu2017PRL,Menke,Shen,Lieu,Zyuzin2018PRB,Yin,Yao1,Yao2,Gong,Shanhui2018PRB,
Bergholtz2018,Rudner2,Kawabata1,Kawabata2,YChen,Haijun2018,YuRui2018,Qibo2019}. In the context of 3D gapless matters with particle gain and loss, Xu and coworkers
introduce a non-Hermitian term in a Weyl Hamiltonian and found that a Weyl point morphs into a Weyl exceptional
ring consisting of exceptional points~\cite{Xu2017PRL}. In contrast
to the nodal ring with a quantized Berry phase and a zero Chern number, the Weyl exceptional ring
possesses both a quantized Berry phase and a quantized (nonzero) Chern number. Remarkably, the Weyl exceptional
ring has recently been experimentally observed in an optical waveguide array~\cite{Rechtsman2018}. Additionally,
when a non-Hermitian term is introduced in a nodal-line semimetal, a nodal ring
develops into two exceptional rings characterized by the Berry phase~\cite{Haijun2018,YuRui2018}. Moreover,
exceptional links and twisted Fermi ribbons were also found in non-Hermitian systems~\cite{Bergholtz2018}.

Compared with solid-state materials, cold atoms provide a clean and highly controllable
platform to simulate Hamiltonians. In the context of topology, there has been
remarkable progress made in cold atoms in both topological insulators and topological gapless matters~\cite{ZollerReview,Zhu2018Rev}.
For topological insulators, the Zak phase was observed in a 1D dimerized optical lattice~\cite{Bloch2013NP}, and the Haldane model, the Chern band and the Harper Hamiltonian
were experimentally realized in cold atoms~\cite{Bloch2013PRL,Ketterle2013PRL,Jotzu2014Nature,Aidelsburger2015Nat,Sengstock2016,Xiongjun2015Sci,Xiongjun2018PRL}.
Long-sought after Thouless pump and the 4D quantum Hall effect were also experimentally observed with cold atoms~\cite{Nakajima2016,Bloch2016,Bloch2018Nature}.
For topological gapless phenomena, the honeycomb
lattices were engineered using laser beams and the Dirac cone was observed by Landau-Zener tunneling~\cite{Sengstock2011Nat,Esslinger2012nature}. The
2D Dirac point was also realized in the Harper Hamiltonian with the flux per plaquette being
1/2 using laser-assisted tunneling~\cite{Bloch2013PRL,Ketterle2013PRL}. Such a point
was also implemented in the hyperfine level space of atoms as the 2D spin-orbit coupling~\cite{Jing2016NP,Jing2016PRL,Xiongjun2015Sci,ShuaiChen2018SB}. Furthermore,
the second Chern number of the Yang monopole has recently been observed in cold atoms~\cite{Spielman2018Science}.

For topological gapless matters in 3D cold atom systems, remarkable experimental and theoretical progress
has been made. In particular, the Weyl nodal line has recently been implemented in the $^{173}$Yb Fermi gas~\cite{Song2018}.
In theory, there has been a number of systems discovered to host topological gapless bands. Specifically,
in superfluids, the Weyl points were found to exist in the quasiparticle energy spectrum of spin-orbit
coupled BCS and Fulde-Ferrell superfluids~\cite{Gong2011prl,Xu2014PRL,Xu2015PRL,Huihu2014PRA} owing to the effective p-wave pairing similar to the $^3$He A
phase. The gapless topological Fulde-Ferrell superfluids with structured Weyl points~\cite{Xu2014PRL,Xu2015PRL,Huihu2014PRA} were also found. Since the dominant pairing is the $p_x+ip_y$ pairing for the attractive
dipole-dipole interacted atoms, the Weyl points can also emerge in the quasiparticle energy spectrum of such
superfluids~\cite{Liubo2015PRL}. When the one-dimensional (1D) spin-orbit coupling is considered,
Weyl nodal rings~\cite{Melo2012PRL} and structured Weyl (type-II)
nodal rings~\cite{Xu2015PRL} were found in the Fermi superfluids.
Recent experimental realization of 1D and 2D spin-orbit coupling in cold atoms~\cite{Spielman2011Nature,JYZhang2012PRL,Jing2012PRL,Zwierlein2012PRL,CQu2013PRA,Olson2014PRA,Luo2014SR,Lev2016PRX,Jo2016PRA,
Jo2016PRA,JLi2016PRL,Ye2017nature,JRLi2017Nature,Song2018SciAdv,Jing2016NP,Jing2016PRL,Xiongjun2015Sci,Xiongjun2018PRL,Song2018} (see review~\cite{Spielman2013Nature,HuiZhai2015,XJLiu2018Review})
has laid the foundation for the experimental observation of Weyl points and nodal rings in superfluids.
Furthermore, Weyl fermion quasiparticles were shown to emerge in a 3D system of polar particles in
magnetic fields~\cite{Ray2015NC} and in a topological density wave phase in cold atomic Rydberg-dressed
atomic fermions~\cite{Xiaopeng2015NC}. Similar to the fermionic excitations, Weyl points and nodal rings were also shown to
emerge in Bogoliubov excitations of bosonic superfluid and Mott insulator phases when
bosonic atoms are loaded into a Weyl semimetal or nodal ring optical lattice~\cite{SPKou2017PRA,HuangYang2018JPB,SPKou2018CPB}.

Since non-interacting systems are easy to understand and control, another direction in cold atoms
is to investigate the experimental scheme for realizing topological gapless bands in non-interacting
systems. This is achievable as interactions can be readily tuned by Feshbach resonances in
cold atoms~\cite{Chin2010RMP}. One can also study the effects of interactions on gapless bands when
the attractive or repulsive interactions are turned on. To date, there have been a number of proposals
for realizing Weyl points and nodal rings. In 2012, Anderson and coworkers
reported an experimental scheme to realize the 3D Weyl spin-orbit coupling by coupling
four hyperfine levels of atoms with Raman laser beams~\cite{Spielman2012PRL}. This scheme can actually achieve a single
Weyl point in the continuous space without lattices. In 2015, based on their experimental realization of a Dirac point in a 2D Harper Hamiltonian,
Dub\v{c}ek and coworkers found an experimental scheme
to realize a Weyl semimetal in a sublattice pseudospin space with laser-assisted tunnelling~\cite{Ketterle2015PRL}.
A year later, based on an experimental realization of the Chern band~\cite{Xiongjun2015Sci}, Xu and Duan
proposed an experimental setup for realizing a Weyl metal~\cite{Yong2016PRATypeII}. In this scheme,
both traditional (type-I) and structured (type-II) Weyl points can be realized and their
transition can be readily tuned by controlling a two-photon detuning. Since the Weyl point is
achieved in the hyperfine level space, the scheme also realizes the 3D Weyl spin-orbit coupling.
Very recently, an experimental scheme to realize a 3D Dirac semimetal in cold atoms
has been proposed~\cite{YongHu2018}.
There are also experimental schemes for achieving the Dirac and Weyl nodal ring semimetal
with cold atoms~\cite{Yong2016DiracRing,DWZhang2016}. Other approaches for constructing Weyl semimetals with cold atoms
include stacking 2D lattice layers with checkerboard-patterned staggered fluxes~\cite{JianHua2012PRA},
stacking one-dimensional Aubry-Andre-Harper models~\cite{sarma2015PRB}, and
shaking
a face-centered-cubic optical lattice~\cite{Lijun2018PRB}.

As a highly controllable system, cold atoms allows us to achieve the manipulation of the location and
number of Weyl points. Such a manipulation changes the anomalous Hall response accordingly. Despite the existence of an inevitable non-adiabatic process when a Hamiltonian parameter is slowly modified, the equilibrated Hall response can be achieved by applying a suitable electric field initially in a coherent dynamics~\cite{YongHu2018}. In addition, cold atoms can be utilized to realize the dynamical Weyl points
and dynamical 4D Weyl nodal rings where one parameter is taken as time~\cite{Yanbin2018PRB}. Such dynamical gapless
phenomena lead to continuous tunable Thouless pumping in higher dimensions.

There are many excellent reviews~\cite{ Turner2013,Hosur2013,Vafek2014,Kharzeev2014,Wehling2014,Witczak-Krempa2014,Burkov2015,Schnyder2015,Hasan2015,Witten2015,
Weng2016,Bansil2016,Burkov2016,Jia2016,Yan2017,Hasan2017,smejkal2017, Burkov2017,Syzranov2018,Armitage2018RMP} on the topic mainly related to topological gapless matters in solid-state
materials. A review mainly focused on topological insulators in cold atoms~\cite{Zhu2018Rev} has appeared recently. In this review,
we summarize the basic concepts of Weyl points, Dirac points, nodal rings and Weyl exceptional
rings, and their presence in quasiparticle spectra of Fermi superfluids and in
single particle spectra of non-interacting Fermi gases. Specifically, in Sec. 2, we introduce the concepts of Weyl points,
Dirac points, nodal rings and Weyl exceptional rings. In Sec. 3, we describe Weyl points and
nodal rings in Fermi superfluids. In Sec. 4, we summarize theoretical proposals for realizing these
topological gapless matters in Fermi gases. In Sec. 5, we briefly summarize other topological
gapless phenomena in cold atoms. Finally, a conclusion and perspective is given.

\section{Gapless points and rings}
In this section, we review several typical topological gapless matters in 3D including
Weyl points, structured (type-II) Weyl points, Dirac points, type-II Dirac points, nodal
rings, structured (type-II) nodal rings, and Weyl exceptional rings.

\subsection{Weyl points}
\begin{figure}[t]
\includegraphics[width=3.2in]{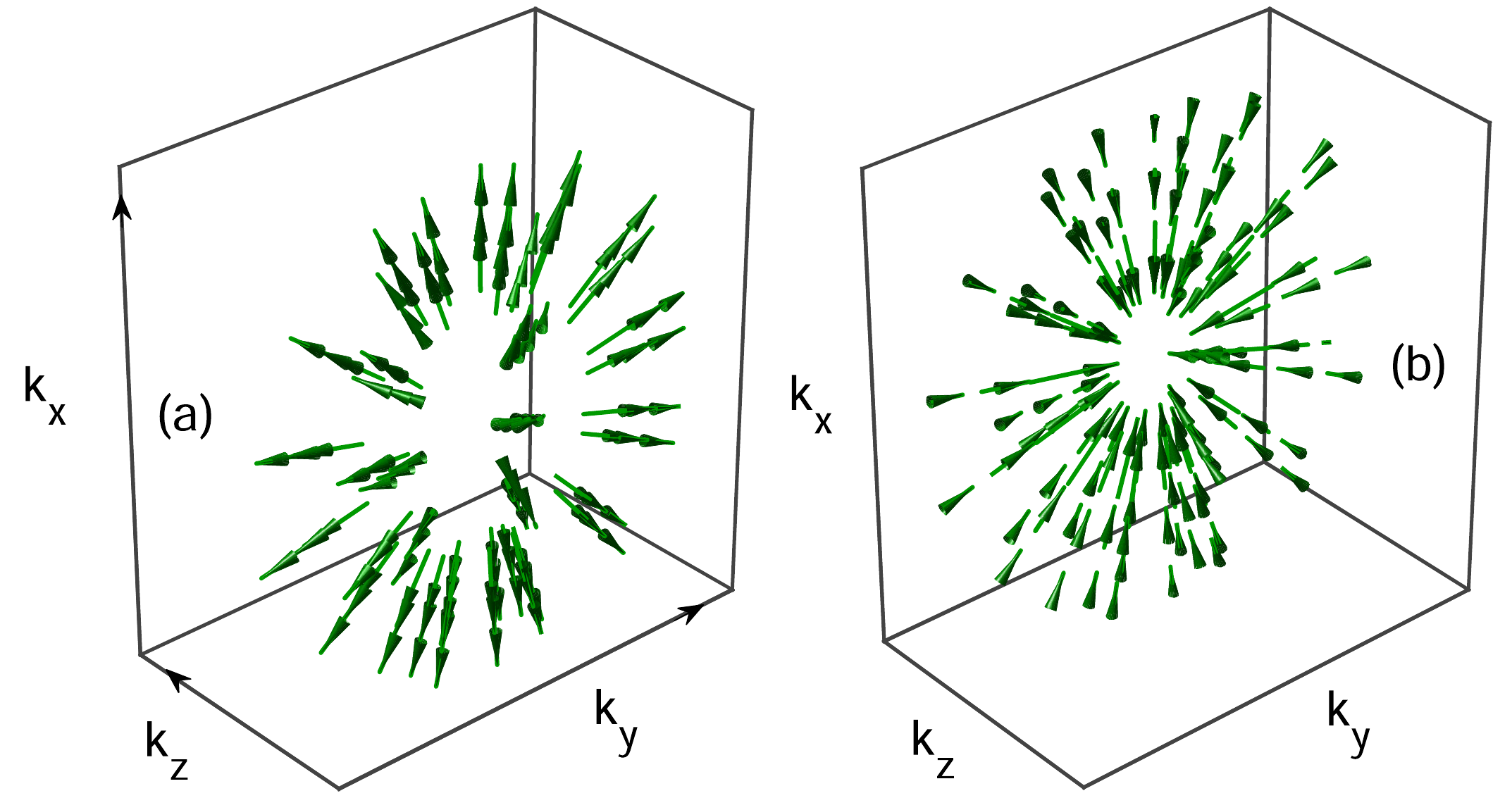}
\caption{(Color online) Schematics of the Berry curvature ${\bm \Omega}_{\mp}({\bf k})$ in (a) and (b), respectively.
The arrows on the axes show their orientation.}
\label{BerryLR}
\end{figure}

In a $C^2$ space, let us consider the following Hamiltonian in momentum space
\begin{equation}
H_W=\epsilon_0({\bf k})\sigma_0 + v{\bm k}\cdot{\bm \sigma}
\label{HW1}
\end{equation}
with the Pauli matrices $\sigma_\nu$ ($\nu=x,y,z$), the $2\times 2$ identity matrix $\sigma_0$, the momentum $k_\nu$ and the real parameter $v$ and $\epsilon_0({\bf k})$, which may be a function of momentum ${\bf k}$. The eigenenergy is $E_\pm ({\bf k})=\epsilon_0({\bf k})\pm  v k$ with $k=\sqrt{k_x^2+k_y^2+k_z^2}$, showing that a double
degeneracy occurs at ${\bf k}=0$. In particle physics, $\epsilon_0({\bf k})$ vanishes and $v=c$ where
$c$ is the light velocity in vacuum, required by the Lorentz symmetry. Apparently, the degenerate point is stable against any perturbation because all $2\times 2$ matrices can be generated by $\sigma_0$, $\sigma_x$, $\sigma_y$ and $\sigma_z$ and
adding any one can at most change the location of the degenerate point instead of gapping it.
The dispersion is linear around all directions and the slope indicates the group velocity. The Weyl fermoins are chiral massless Dirac fermions whose velocities are parallel or anti-parallel to their spins. In addition, the Weyl point
can be regarded as a monopole of Berry curvatures [see Fig.~\ref{BerryLR}] in momentum space given by
\begin{equation}
{\bm \Omega}_{\pm}({\bf k})=\mp \frac{\bf k}{2k^3},
\end{equation}
where
\begin{equation}
{\bm \Omega}_{\pm}({\bf k})=i\langle \nabla u_{\pm}({\bf k})|\times|\nabla u_{\pm}({\bf k})\rangle
\end{equation}
is the Berry curvature
with $|u_{\pm}({\bf k})\rangle$ being the upper and lower eigenstate of the Weyl Hamiltonian so that $H_W|u_{\pm}({\bf k})\rangle=E_{\pm}({\bf k})|u_{\pm}({\bf k})\rangle$.
The topology of the Weyl point is characterized by the integral of the Berry curvature over a closed surface $S$
in momentum space enclosing the point, i.e.,
\begin{equation}
C_{\pm}=\frac{1}{2\pi} \oint_S d{\bf S}\cdot {\bm \Omega}_{\pm}({\bf k}),
\end{equation}
where the Chern number $C_{\pm}$ is defined and hence $C_{\pm}={\mp}1$, implying that the system is a
topological insulator on this closed surface.

In a continuous space without lattices, a single Weyl point can appear in band structures.
This may occur in cold atom systems, where optical lattices are naturally absent.
In fact, Anderson and coworkers proposed an experimental scheme in ultracold atomic
gases to realize a single Weyl point~\cite{Spielman2012PRL}. Since the $C^2$ space corresponds to the hyperfine level space,
the Weyl point also corresponds to the 3D Weyl spin-orbit coupling.
In contrast, in the presence of lattices,
which is generally the case in solid materials, Weyl points have to appear in pairs so that their total Chern number
vanishes. Otherwise, let us choose a closed surface wrapping all Weyl points (the Chern number on this surface
is nonzero) and
continuously enlarge the surface so that it shrinks to a small sphere
as the surface travels across Brillouin boundaries. On the infinitesimally small sphere,
the Berry curvature does not diverge as the system on the sphere is always gapped and the gap does not vanish when
the sphere becomes smaller. In this case, the integral of Berry curvatures on the sphere vanishes.
However, this cannot happen, since
the system on the surface is always gapped due to the absence of other gapless points and
the Chern number on this surface should remain unchanged.

Alternatively, we may regard a Weyl point as the phase transition point between a Chern insulator and
a topologically trivial insulator in momentum space in terms of a parameter, say $k_z$. In the
parameter region for $k_z$ where the insulator is topological, there are chiral edge states in the
presence of edges along other directions perpendicular to the $z$ axis. Because these edge states appear only in topological insulator
region for $k_z$, but does not in the topologically trivial insulator region, we obtain the Fermi surface consisting of
chiral edge modes, which connects a pair of Weyl points with opposite charges, if the Fermi surface is located at
the Weyl point; these surface states form the Fermi arc, which is not closed on a surface.
In superfluids or superconductors, the surface states correspond to the chiral Majorana modes~\cite{Balent2012PRB}.

\begin{figure}[t]
\includegraphics[width=3.2in]{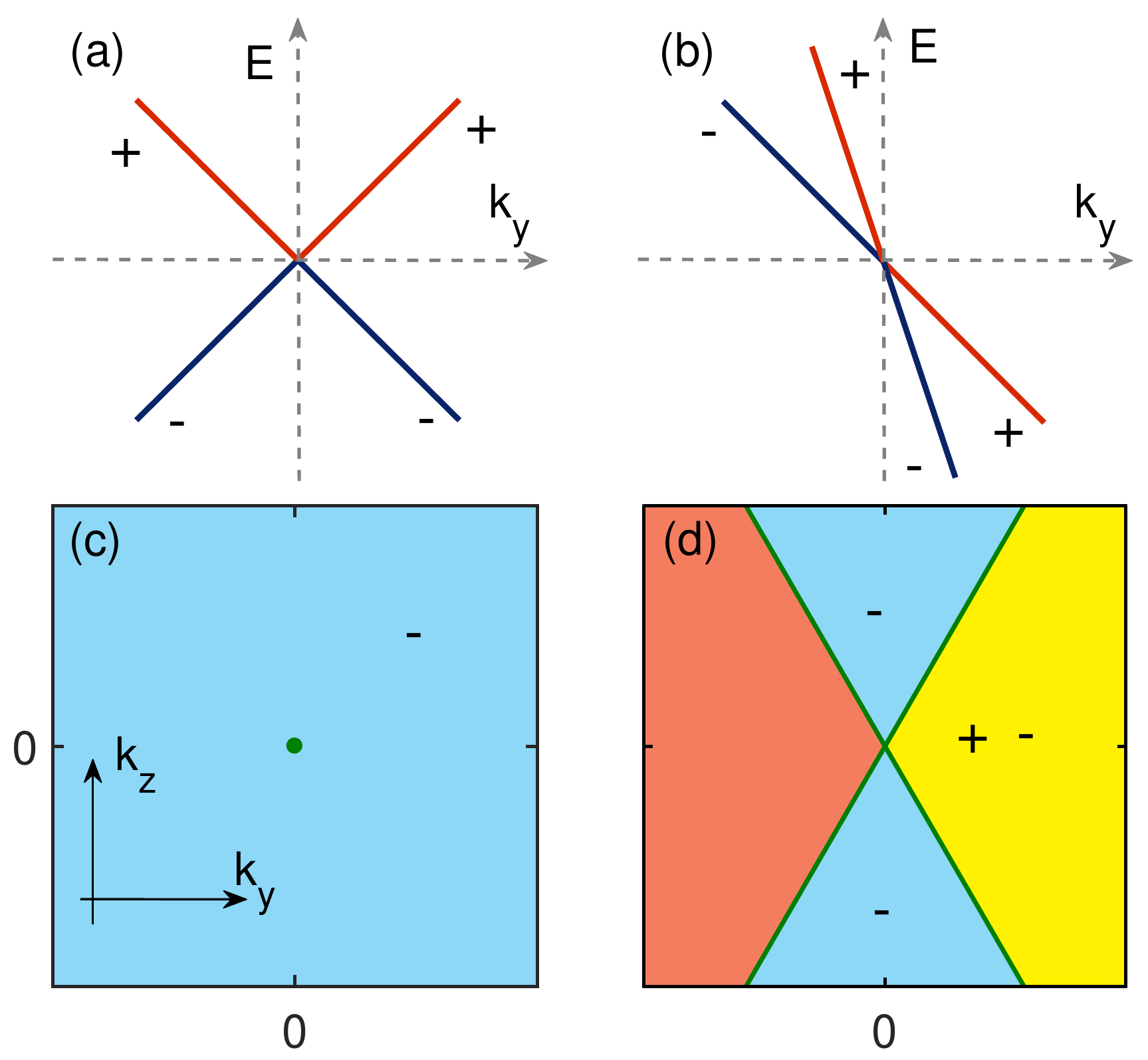}
\caption{(Color online) Energy bands with respect to $k_y$ ($k_x=k_z=0$) for a traditional (type-I) Weyl point (a)
and a structured (type-II) Weyl point (b). Profiles of band occupation in the $k_x=0$ plane for a traditional (type-I)
Weyl point (c) with a point Fermi surface (denoted by a green circle) and a structured (type-II) Weyl point (d) with an open Fermi surface (denoted by two green lines). In (a-b), $\pm$ label the eigenvalues of $\sum_{i}k_{i}\sigma _{i}/k$, i.e., the
helicity. In (c-d), $\pm$ represent the occupied band with the corresponding helicities. In the red region, no bands are occupied.
Adapted from Ref. [\cite{Xu2015PRL}].}
\label{StructuredWP}
\end{figure}

In general, the existence of Weyl points does not require any symmetry. But materials
usually have some symmetries which may have some constraints on Weyl semimetals. For instance, if a system possesses
both time-reversal $\mathcal{T}$ (i.e., $\mathcal{T}H({\bf k})\mathcal{T}^{-1}=H(-{\bf k})$) and inversion symmetry $\mathcal{P}$ (i.e., $\mathcal{P}H({\bf k})\mathcal{P}^{-1}=H(-{\bf k})$) or the PT symmetry
$\Theta_{\text{PT}}=\mathcal{P}\mathcal{T}$ (i.e., $\Theta_{\text{PT}}H({\bf k})\Theta_{\text{PT}}^{-1}=H({\bf k})$),
the band is guaranteed to be doubly degenerate when $\Theta_{\text{PT}}^2=-1$. This arises
from the fact that if $|u({\bf k})\rangle$ is an eigenstate with energy being $E_{\bf k}$,
$\Theta_{\text{PT}}|u({\bf k})\rangle$ is the other distinct eigenstate with the same energy,
since if they correspond to the identical state, i.e., $\Theta_{\text{PT}}|u({\bf k})\rangle=e^{i\lambda}|u({\bf k})\rangle$,
then $(\Theta_{\text{PT}})^2|u({\bf k})\rangle=|u({\bf k})\rangle=(-1)|u({\bf k})\rangle$,
which is impossible.
In this case, a touching point, if exists, has to be a fourfold degenerate point,
which, as will be shown later, is a Dirac point with a vanishing Chern number rather than a
Weyl point with a nonzero Chern number. On the other hand, a band is generally not degenerate
if $\Theta_{PT}^2=1$. Yet, in this case, we have ${\bm \Omega}_n({\bf k})=0$ if both time-reversal
and inversion symmetries are preserved. Overall, to achieve a Weyl point in
band structures, one has to break either
time-reversal or inversion symmetry or both. With only time-reversal symmetry,
we have
${\bm \Omega}_n(-{\bf k})=-{\bm \Omega}_n({\bf k})$ and hence if there is a Weyl point at
${\bf k}={\bf k}_W$ with the Chern number being $C$ on a closed surface enclosing it,
there must exist another Weyl point at ${\bf k}=-{\bf k}_W$ with the same Chern number $C$~\cite{Halasz2012PRB}.
This indicates that there must be at least two pairs of Weyl points in a time-reversal
Weyl semimetal. With only inversion symmetry, we have ${\bm \Omega}_n(-{\bf k})={\bm \Omega}_n({\bf k})$
and then if there exist a Weyl point at ${\bf k}={\bf k}_W$, there must exist another one at ${\bf k}=-{\bf k}_W$
with the opposite Chern number~\cite{Wan2011prb,Burkov2011PRL}.
In this case, we can have a pair of Weyl points in the system.
When a system has neither time-reversal and inversion symmetry nor PT symmetry,
a pair of Weyl points can also appear. Many cold atom systems
belong to this category, such as in the spin-orbit coupled superfluids~\cite{Gong2011prl,Xu2014PRL,Xu2015PRL}.

In particle physics, $\epsilon_0({\bf k})$ cannot appear due to the Lorentz symmetry. However,
in condensed matter systems including ultracold atomic gases, the symmetry is not necessarily
respected as the energy of a particle is much smaller than their static masses. As a consequence,
this term can emerge. If this term is momentum dependent, e.g.,
\begin{equation}
H_{SWP}=-\alpha k_y \sigma_0+{\bf k}\cdot {\bm \sigma},
\label{SWToy}
\end{equation}
with the eigenenergy being $E=-\alpha k_y \pm k$.
Clearly, the Weyl cone is tilted when $0<|\alpha|<1$. This tilt has also been studied in the graphene~\cite{Goerbig2008PRB}.
More interestingly, when $|\alpha|>1$, the Fermi surface becomes open (as shown in Fig.~\ref{StructuredWP}(b)) and the density of states
is finite instead of zero for the traditional Weyl point~\cite{Xu2015PRL,Soluyanov2015Nature}.
This new phenomenon was initially discovered in the quasiparticle spectrum of the gapless topological Fulde-Ferrell superfluids in spin-orbit coupled gases with attractive interactions~\cite{Xu2014PRL}. Later,
the effective Hamiltonian was proposed to describe it, and the new Weyl point was coined structured Weyl points~\cite{Xu2015PRL}.
In superfluids, since a structured Weyl point is located at zero energy, the other has to be at the
same energy due to the intrinsic particle-hole symmetry.
These new fermions were also predicted in band structures of solid-state materials, which are normally located at distinct energies,
and were named type-II Weyl points~\cite{Soluyanov2015Nature}.
Despite the presence of the tilt term, the topology of systems
with structured (type-II) Weyl points is still fully determined by the latter
terms, leading to the same Fermi arc connecting two Weyl points.

Above we mainly focus on the Weyl point with the Chern number $C=\pm 1$. In fact, it can be readily
generalized to a degenerate point with higher Chern numbers~\cite{Bernevig2012PRL}, which is described by the following Hamiltonian
\begin{equation}
H_{MW}=\left(
    \begin{array}{cc}
      k_z & k_\perp^n e^{-i n \phi} \\
      c.c. & -k_z \\
    \end{array}
  \right)
\label{MWHam}
\end{equation}
with $n$ being a positive integer, $k_\perp=\sqrt{k_x^2+k_y^2}$, $k_x=k_\perp \cos(\phi)$ and $k_y=k_\perp \sin(\phi)$.
The eigenenergy is $E_{\pm}=\pm\sqrt{k_z^2+k_\perp^{2n}}$.
The Berry curvatures corresponding to the two bands are ${\bm \Omega}_{\pm}({\bf k})=\mp \frac{nk_\perp^{2n-1}}{2E_+^3}({\bf e}_\perp +\frac{nk_z}{k_\perp}{\bf e}_z)$ and their integral over a closed surface wrapping the degenerate point gives the Chern number $\mp n$. Additionally, the dispersion is no longer linear
along $k_x$ and $k_y$ directions when $n>1$. For instance, when $n=2$, the dispersions along
these directions are quadratic. These degenerate points are called multiple Weyl points and, specifically,
double Weyl points when $n=2$~\cite{Bernevig2012PRL}. Multiple Weyl points with topological charges being 2 and 3 have
recently been experimentally observed in photonic crystals~\cite{Wenjie2016nc}. For double Weyl points, tight-binding Hamiltonians in a 3D cubic optical lattice was constructed~\cite{Lepori2016PRA,DanWei2017PRA} and an
experimental scheme based on the laser-assisted tunneling with cold atoms was put forward~\cite{DanWei2017PRA}.

\subsection{Dirac points}
A Dirac point in 3D refers to a fourfold degeneracy with the zero Chern number, i.e.,
a Dirac point consists of two Weyl points with the opposite topological charges located at the same momentum. It is
described by the Hamiltonian
\begin{equation}
H_D=v \tau_z {\bf k}\cdot{\bm \sigma},
\end{equation}
where $\tau_z$ is also a Pauli matrix. $\bm \tau$ and $\bm \sigma$ may refer to the orbital and
spin degrees of freedom. Since the Dirac point corresponds to
a massless Dirac fermion dictated by the Dirac equation, it is fundamentally important.

In comparison with the Weyl point that is stable against any small perturbations, a Dirac point
is unstable against perturbations. The reason is that a Dirac point is depicted by a $4\times4$
Hamiltonian, which to be totally determined requires 16 matrices. Since only three matrices
are used here, other matrices, e.g., $\tau_x$, may open a gap if exist. Therefore, in
solid materials, symmetries are required to exclude these additional terms that can open a gap. For a system with both time-reversal and inversion symmetry so that $\Theta_{PT}^2=-1$,
both valence and conductance bands are doubly degenerate and ${\bm \Omega}_{n_1}({\bf k})=-{\bm \Omega}_{n_2}({\bf k})$
corresponding to two degenerate bands labeled by $n_1$ and $n_2$.
When these bands have
a degeneracy with linear dispersions along all three directions, it must be a Dirac point with vanishing Chern numbers.
However, even with these symmetries, the Dirac points are still not guaranteed to exist. They can either appear accidentally, for example,
as a critical point through the normal insulator to $Z_2$ topological insulator,
or appear due to space symmetries, such as additional uniaxial rotational symmetry~\cite{Nagaosa2014NatCom}.

In cold atom systems, the philosophy is different. Here, a Hamiltonian is simulated by exquisitely
controlling laser beams. From this perspective, to implement a Dirac point, one needs to conceive
a concrete continuous Hamiltonian and analyze the symmetry of the Hamiltonian to see whether a Dirac
point can appear or not, instead of writing down a Hamiltonian through analyzing a system's symmetry.
Xu and Hu recently proposed a Hamiltonian~\cite{YongHu2018}
\begin{equation}\label{eq:Hc}
H_{C}=\frac{{\bf p}^{2}}{2m}-\sum _{\nu=x,y,z}V_{\nu}\cos^{2}(r_{\nu}\pi/a)+m_{z}\sigma_{z}+V_{\textrm{SO}},
\end{equation}
with the mass of atoms $m$, the momentum operator ${\bf p}$, $V_{\textrm{SO}}=M_{y}\sigma_x-M_{x}\sigma_y$ depicting the nondiagonal optical lattices with $M_{x}=\Omega_{SO}\sin(r_{x}\pi/a)\cos(r_{y}\pi/a)\cos(r_{z}\pi/a)$ and
$M_{y}=\Omega_{\textrm{SO}}\sin(r_{y}\pi/a)\cos(r_{x}\pi/a)\cos(r_{z}\pi/a)$. An experimental realization setup
in cold atoms is also proposed. When $m_z=0$, this Hamiltonian respects both time-reversal and inversion
symmetry, the energy band is doubly degenerate. Interestingly, the band structure exhibits two Dirac
points at $(k_x,k_y,k_z)=(0,\pi,\pm\pi/2)$. With nonzero $m_z$, the time-reversal symmetry is broken and each Dirac
point will split into two Weyl points.

A Dirac point can also become type-II. Its type-II counterpart has two forms:
both points (as a Dirac point can be viewed as consisting of two Weyl points)
are type-II, and one is type-I and the other type-II. The former happens when
we add a term $-\alpha k_y \sigma_0$ wiht $|\alpha|>1$ into the Hamiltonian.
The latter hybridized structure appears when we include a term
\begin{equation}
\left(
                                                                               \begin{array}{cc}
                                                                                 -\alpha k_y \sigma_0 & 0 \\
                                                                                 0 & 0 \\
                                                                               \end{array}
\right) \text{or}
\left(
                                                                               \begin{array}{cc}
                                                                                 0 & 0 \\
                                                                                 0 & -\alpha k_y \sigma_0 \\
                                                                               \end{array}
\right)
\end{equation}
with $|\alpha|>1$.
Type II Dirac points have been theoretically predicted and experimentally observed in solid
materials~\cite{Huang2016PRB,YanNC2017,Christodoulides2017PRL,JianHua2017,Park2017,Guanghou2017}.

\subsection{Nodal rings}
Before we elaborate on the Nodal ring in 3D, let us first discuss the Dirac point in 2D.
It is important to note that for the Dirac equation in 2D, since the $\gamma$ matrices
are $2\times 2$, there are no Weyl fermions.
Similar to Weyl points in 3D, a Dirac point in 2D also exhibits the
linear dispersion along all two directions and can be described by
\begin{equation}
H_{2D}=\epsilon_0({\bf k})+k_x\sigma_x+k_y\sigma_y.
\label{H2DDirac}
\end{equation}
A celebrated system hosting Dirac points is the graphene with a honeycomb lattice structure~\cite{Geim2009RMP,DasSarma2011RMP}.
In contrast to the Weyl point which is stable against perturbations, the 2D Dirac point is unstable. For example,
the presence of a small $m\sigma_z$ term can open a gap. Therefore, similar to the 3D Dirac point,
a symmetry is required to protect it.
One may consider two types of symmetries protecting a 2D Dirac point: chiral symmetry and $PT$ symmetry.
For the former, one can define the $Z$ topological invariant: the winding number to characterize
the degenerate point. For instance, for the Hamiltonian (\ref{H2DDirac}) with the chiral
symmetry, i.e., $\sigma_zH({\bf k})\sigma_z=-H({\bf k})$,
both $\sigma_z$ and $\epsilon_0$ term are not allowed in the Hamiltonian~(\ref{H2DDirac}) so that the gap cannot be opened.
There, the winding number $W$ can be defined as the number of times that $(k_x,k_y)$ wraps around the origin
while it varies along a path enclosing the degenerate point in momentum space. Note that the corresponding Berry phase (or Zak phase along the path) is $\gamma(C)=W\pi$.
However, the chiral symmetry is generally not an exact symmetry of band structures. In addition, breaking the chiral symmetry
may not gap the point inevitably.
For example, in the Hamiltonian (\ref{H2DDirac}), while $\epsilon_0(k)$ breaks the chiral symmetry,
it cannot open a gap. With $\epsilon_0(k)$, while the winding number is not well defined,
the Berry phase is still well defined and quantized.
In fact, we can consider another symmetry, that is, the $PT$ symmetry~\cite{Kim2015PRL,ZDWang2016PRL}
with $\Theta_{PT}^2=1$. Under this symmetry, we can define a $Z_2$ topological invariant (instead of $Z$)~\cite{Kim2015PRL,ZDWang2016PRL}
$\omega(C)=e^{i\gamma(C)}=\pm 1$ corresponding to a quantized Berry phase along a closed path in
momentum space. Here, the $PT$ symmetry ensures that
$\omega(C)=\pm 1$. For the Hamiltonian (\ref{H2DDirac}), $\Theta_{PT}=\sigma_x \mathcal{K}$ such that
$\Theta_{PT}^{-1}H({\bf k})\Theta_{PT}=H({\bf k})$. It is easily seen that $\epsilon_0$ is allowed while
$\sigma_z$ is forbidden, preventing the gap from being opened.

Let us extend the Dirac point in 2D to 3D such that the degenerate point
develops into a line or a closed ring. Specifically, let us consider a simple toy model described
by the Hamiltonian,
\begin{equation}
H_{WN}=\epsilon_0({\bf k})+k_x\sigma_x+(k^2-m^2)\sigma_y,
\label{HNodalRing}
\end{equation}
where $k^2=k_x^2+k_y^2+k_z^2$ and $m$ denotes a nonzero mass term. The eigenenergy is $E_{\bf k}=\epsilon_0({\bf k})\pm \sqrt{k_x^2+(k^2-m^2)^2}$,
leading to a degenerate ring appearing at $k_x=0$ and $k_y^2+k_z^2=m^2$. Although the degenerate ring cannot
be described by the Dirac equation or Weyl equation in 3D, we may follow the convention to call it
Weyl nodal ring given that it is doubly degenerate.
To see the connection of the Weyl nodal ring to the Dirac point in 2D, let us fix
$|k_z|<|m|$ and obtain two gapless points located at $k_{Dx}=0$
and $k_{Dy}=\pm m_0$ with $m_0=\sqrt{m^2-k_z^2}$. In the vicinity of these points, the Hamiltonian is given by
$H=\epsilon_0({\bf k})+k_x\sigma_x \pm 2m_0\delta k_y \sigma_y$ to the first order, where $\delta k_y$ is measured with respect to $k_y=\pm m_0$; evidently, this Hamiltonian depicts the 2D Dirac points.

Analogous to the 2D Dirac point, the Weyl nodal ring is unstable and a small perturbation of $\sigma_z$ can open
a gap for the degeneracy. Hence symmetry is required to protect them, ensuring the absence of $\sigma_z$ in the
Hamiltonian (\ref{HNodalRing}). Likewise, let us consider two symmetries: chiral symmetry and $PT$ symmetry. The chiral symmetry excludes $\epsilon_0$ and $\sigma_z$ terms and the
winding number can be defined over a closed trajectory enclosing the ring. As we have noted, the chiral symmetry
is generally not an exact symmetry because $\epsilon_0$ is generally nonzero. In this case, we may consider
the $PT$ symmetry, which ensures that $\sigma_z$ cannot exist while $\epsilon_0$ can. In this case, a $Z_2$
topological invariant
$\omega(C)=e^{i\gamma(C)}=\pm 1$ can be defined over a closed path enclosing the ring to characterize
it~\cite{Kim2015PRL}.

To see whether surface states appear on the boundaries in the Weyl nodal system, let us
consider a semi-infinite system located at $x>0$ and the open boundary is imposed at $x=0$.
The system can be described by the effective Hamiltonian in the real space along the $x$ direction
\begin{equation}
H=-i\partial_x \sigma_x+(-\partial_x^2+M)\sigma_y.
\end{equation}
Solving the time independent Schr\"odinger equation gives us a zero energy solution $\psi=\phi(x)|\uparrow\rangle$ with $\phi(x)\sim e^{-\lambda_+ x}-e^{-\lambda_- x}$ with $\lambda_{\pm}=\frac{1}{2}(1 \pm \sqrt{1+4M})$
and $M<0$; the solution is locally localized around $x=0$ and thus corresponds to the surface state~\cite{ShunqingBook}.
For the Hamiltonian (\ref{HNodalRing}), $M=k_y^2+k_z^2-m_0^2$ and, hence, there appear
zero energy surface states, when $k_y^2+k_z^2<m_0^2$. Since these surface states have zero energy, they provide another platform to study strongly correlated physics, such
as high-temperature superconductivity~\cite{Volovik2011}. In the presence
of $\epsilon_0(k_y,k_z)$, the energy dispersions of these surface states become $\epsilon_0(k_y,k_z)$,
showing that the flat band obtains dispersion and becomes drumhead structure.
In superfluids or superconductors, the surface states are Majorana zero modes.

The degenerate ring may be fourfold degenerate, which is described by the Hamiltonian
\begin{equation}
H_{DN}=k_x\sigma_x+(k^2-m^2)\tau_z\sigma_y.
\label{HDiracNodalRing}
\end{equation}
This Hamiltonian possesses both
$\tau_z$ (i.e., $\tau_z H \tau_z=H$) and chiral symmetries (i.e., $\sigma_z H \sigma_z=-H$).
Hence the ring can be characterized by a pair of winding numbers $(W_+,W_-)$ defined in the
subspaces corresponding to $\tau_z=\pm 1$ over a closed path enclosing the gapless
points on the ring. For the Hamiltonian (\ref{HDiracNodalRing}), the pair of winding
numbers is $(1,-1)$. Similar to a 3D Dirac point with a pair of Chern numbers being
$\pm(1,-1)$, the nodal ring is called Dirac nodal ring.

In solid-state materials, the energy spectrum is generally very complicated and
a nodal ring is generally not located at the same energy.
In cold atom systems, Xu and Zhang proposed a Hamiltonian with its experimental realization
scheme~\cite{Yong2016DiracRing}, illustrating the presence of Weyl and Dirac nodal rings that almost lie
at the same energy. This Hamiltonian in the real space is
\begin{equation}
H_{CN}=\frac{\mathbf{p}^{2}}{2m}-\sum_{\nu =x,y,z}V_{\nu }\cos ^{2}(r_{\nu }\pi/a_\nu)+h_{z}\sigma _{z}-V_{SO}\sigma _{y},
\label{NodalLineH}
\end{equation}%
with $V_{SO}=\Omega _{SO}\sin (r_{x}\pi/a_x)\cos (r_{y}\pi/a_y)\cos (r_{z}\pi/a_z)$ and
$V_{\nu }$ and $\Omega _{SO}$ characterizing the strength of optical lattices.
The Hamiltonian in momentum space of its tight-binding counterpart is given as
\begin{equation}
H_{CN}(\mathbf{k})=-h_{t}\tau _{x}+h_{z}\sigma _{z}+d_{x}\tau _{y} \sigma
_{y},  \label{HTk}
\end{equation}%
where $%
h_{t}=2\sum_{\nu }t_{\nu=x,y,z }\cos (k_{\nu }a)$ and $d_{x}=2t_{SO}\sin
(k_{x}a_x)$ with $t_{\nu}$ and $t_{SO}$ denoting the tunneling strength.
It is easily seen that there is a Dirac nodal ring when $h_z=0$, which
splits into two Weyl nodal rings when $h_z\neq 0$.

In the presence of $\epsilon_0({\bf k})$ that is momentum dependent, for instance,
$\epsilon_0({\bf k})=-\alpha k_x \sigma_0$, the Hamiltonian (\ref{HNodalRing}) leads to a structured nodal ring~\cite{Xu2015PRL} (also called type-II nodal ring)
when $|\alpha|>1$. Such a ring can occur in both quasiparticle spectra of superfluids~\cite{Xu2015PRL} and
band structures of solid materials~\cite{Shengyuan2017PRB,Kou2017}; it has recently
been experimentally realized~\cite{Weiwei2018}.

\subsection{Weyl exceptional rings}
When particle gain and loss, which are generally present in natural systems, are introduced,
a system can be described by a non-Hermitian Hamiltonian~\cite%
{Moiseyev2011, Berry2004, Rotter2009, Heiss2012}. Such a
Hamiltonian generically exhibits complex eigenvalues unless the
$\mathcal{PT}$ symmetry~\cite{Bender1998} holds, and the imaginary
part of such complex eigenvalues is associated with either decay or growth.
In non-Hermitian systems, there exist a special ``degenerate'' point, called
exceptional point, where two eigenstates coalesce and the Hamiltonian becomes defective.

Xu and coworkers introduced a non-Hermitian term in the Weyl Hamiltonian, which is described by the toy model
\begin{equation}
H_{\text{WER}}=i\gamma\sigma_z+{\bf k}\cdot {\bf \sigma},
\label{WERH}
\end{equation}
with $i\gamma\sigma_z$ denoting the non-Hermitian term~\cite{Xu2017PRL}. Here, only the simplest $i\gamma\sigma_z$
is included. One may also incorporate other terms such as $i\gamma_x\sigma_x$ and $i\gamma_y\sigma_y$,
but the Hamiltonian can always be transformed into the form of the Hamiltonian (\ref{WERH}).

\begin{figure*}[t]
\includegraphics[width=\textwidth]{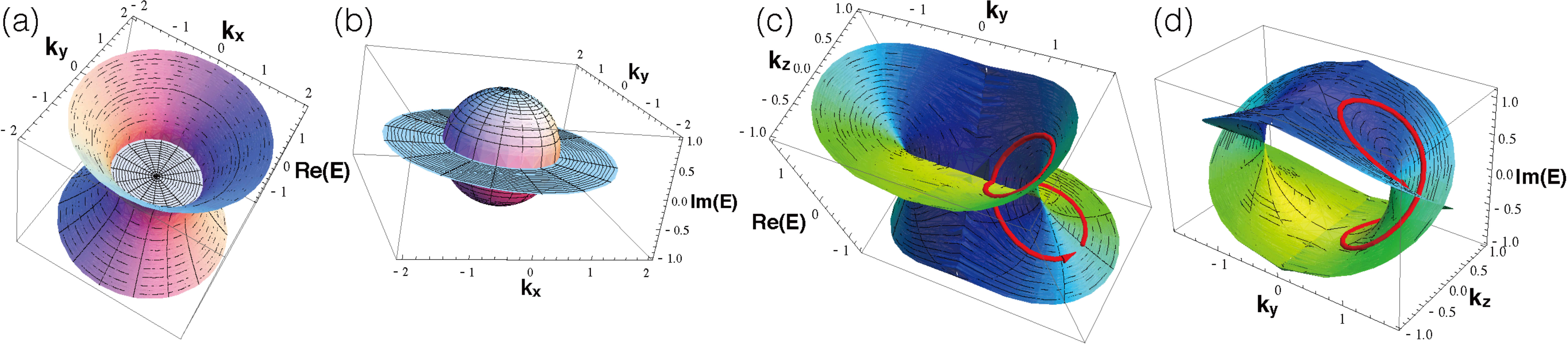}
\caption{(Color online) Real (a) and imaginary parts (b) of energy spectra
in the $(k_x,k_y)$ plane for $%
k_z=0$ . Real (c) and imaginary parts (d) of the Riemann surface with respect to $k_y$ and $k_z$ for $k_x=0$.
The color indicates the strength of $\theta \text{ mod }4\pi $.
Reproduced from Ref. [\cite{Xu2017PRL}].}
\label{WER}
\end{figure*}

The eigenvalues of the Hamiltonian are $E_{\theta }(\mathbf{k}%
)=\sqrt{k^{2}-\gamma ^{2}+2ik_{z}\gamma }=\sqrt{A(\mathbf{k})}e^{i\theta /2}$%
, where $A(\mathbf{k})=\sqrt{(k^{2}-\gamma ^{2})^{2}+4k_{z}^{2}\gamma ^{2}}$
with $k^{2}=k_{x}^{2}+k_{y}^{2}+k_{z}^{2}$, and $\theta $ is defined by $%
\cos \theta =(k^{2}-\gamma ^{2})/A(\mathbf{k})$ and $\sin \theta
=2k_{z}\gamma /A(\mathbf{k})$. Since $e^{i\theta /2}$ obtains a minus sign
upon $\theta\rightarrow \theta +2\pi $, $\theta$ can be used to denote two
energy bands. Without $\gamma$, there is a Weyl point at ${\bf k}=0$ and $\theta$ can take only
two nonequivalent discrete values: $0$ and $2\pi$, corresponding to the two bands.
With $\gamma$, the single touching point develops into a ring
in the $k_z=0$ plane characterized by $k=|\gamma|$; on this ring consisting of
exceptional points where two eigenstates coalesce into a single one, both
real and imaginary parts of energy vanish [see Fig.~\ref{WER}]. In this case,
$\theta$ can take continuous values from $0$ to $4\pi$
and a state in one band ends up with another state in the other band if
$\theta$ travels $2\pi$ across branch points that the exceptional ring plays
the role of.

To characterize the topology of the Weyl exceptional ring, we need
to find a closed surface in momentum space enclosing the ring; on the
surface, a topological invariant can be defined by states $|u_{n}({\bf k})\rangle$,
where $n$ and $\bf k$ denote the band index and momentum. However, for the
Weyl exceptional ring, the energy is multi-valued and we need to
make sure that the energy is continuous on the closed surface.
In complex analysis, we can use the Riemann surface, a two-dimensional manifold
which wraps around the complex plane infinite (noncompact) or finite (compact)
number of times, to describe a multi-valued function [see Fig.~\ref%
{WER}(c) and (d) for the Riemann surface of $E_{\theta }$ for $k_{x}=0$
, on which the energy is single-valued and the surface connects the
different bands]. We, therefore, define the states on the closed surface
living on the Riemann surface, which connects different bands.
For instance, consider a state at $\mathbf{k}_{0}$ with $\theta _{0}$,
we can obtain any other states on
the surface $\mathcal{S}$ by starting from this state and
travelling on momentum space surface $\mathcal{S}$ while keeping $%
E_{\theta }(\mathbf{k)}$ on the Riemann surface.

The topology of a Weyl exceptional ring can be characterized by
both the Chern number and Berry phase. For the former, there are two approaches
to define the Chern number. One is based on the integral of spin vector fields~\cite{KlemanSoft},
that is,
\begin{equation}
N_3=\frac{1}{4\pi}\oint_{\mathcal{S}}\mathbf{d}_\theta \cdot \left(\frac{%
\partial \mathbf{d}_\theta}{\partial u_1}\times\frac{\partial \mathbf{d}%
_\theta} {\partial u_2}\right)du_1 du_2.
\end{equation}
$N_3$ describes the number of times that the spin field $\mathbf{d}%
_\theta=-\sum_{\nu=x,y,z}\langle\sigma_\nu\rangle \mathbf{e}_\nu$ wraps
around a closed surface $\mathcal{S}$ parametrized by ($u_1$,$u_2$) in
momentum space as defined above. Here,
$%
\langle\sigma_\nu\rangle\equiv \langle u_{\theta}(\bm k)|\sigma_\nu|u_{%
\theta}(\bm k)\rangle$ with $|u_{\theta}(\bm k)\rangle$ being the normalized
right eigenstate of $H(\mathbf{k})$ [i.e., $H(\mathbf{k})|u_{\theta}(\bm %
k)\rangle=E_\theta(\mathbf{k})|u_{\theta}(\bm k)\rangle$ and $\langle
u_{\theta}(\bm k)|u_{\theta}(\bm k)\rangle=1$], and
$
\mathbf{e}_\nu$ is the unit vector along the $\nu$ direction.

Another approach to define the first Chern number is based on the Berry curvature
that is,
\begin{eqnarray}
C_2^{RR}&=&\frac{1}{2\pi}\oint_{\mathcal{S}}{\bm \Omega_\theta}^{RR}(\bm k)\cdot d{\bm S}, \\
C_2^{LR}&=&\frac{1}{2\pi}\oint_{\mathcal{S}}{\bm \Omega_\theta}^{LR}(\bm k)\cdot d{\bm S}%
.
\label{ChernEq}
\end{eqnarray}
Here we consider two types of the Berry curvature defined by the right eigvenstate and
the left and right ones. For the former,
${\bm \Omega}_{\theta}^{RR}(\bm k)=i\langle \nabla_{\bm k} u_{\theta}(\bm %
k)|\times |\nabla_{\bm k} u_{\theta}(\bm k)\rangle$. For the latter,
${\bm \Omega}_{\theta}^{LR}(\bm k)=i\langle \nabla_{\bm k} \tilde{u}_{\theta}(\bm %
k)|\times |\nabla_{\bm k} u_{\theta}(\bm k)\rangle=-\frac{\tilde{\bf k}}{2 E_\theta^3({\bf k})}$
with $\tilde{\bf k}=k_x{\bf e}_x+k_y{\bf e}_y+(k_z+i\gamma){\bf e}_z$ where $\langle \tilde{u}_{\theta}(\bm k)|$ is
the normalized left eigenstate of $H$ [i.e., $\langle \tilde{u}_{\theta}(\bm %
k)|H(\mathbf{k})=\langle \tilde{u}_{\theta}(\bm k)|E_\theta(\mathbf{k})$ and
$\langle\tilde{u}_\theta(\mathbf{k})|u_\theta(\mathbf{k})\rangle=1$].
While the local Berry curvatures are distinct for these two definitions,
the Chern number obtained by their integral is equal, that is,
$N_3=C_2^{RR}=C_2^{LR}=\pm 1$ when the surface $\mathcal{S}$
wraps the Weyl exceptional ring and $C_2^{RR}=C_2^{LR}=0$ otherwise.

To understand the physical meaning underlying the Berry curvature,
let us consider a wave packet described by the center coordinate
$\mathbf{r}_c$ and $\mathbf{k}_c$
in the real space and momentum space, respectively.
The dynamics of the wave packet under an external gradient force $%
\mathbf{F}$ is governed by the following the semiclassical equation (see the supplemental material in
Ref. [\cite{Xu2017PRL}] for derivation),
\begin{eqnarray}
\dot {\mathbf{r}}_{c}&=&\partial_{\mathbf{k}_{c}}\bar{E}({\mathbf{k}_{c}})-%
\dot {\mathbf{k}}_{c}\times{\bm \Omega_\theta}({\bm k}_c), \\
\hbar\dot {\mathbf{k}}_{c}&=&\mathbf{F},
\end{eqnarray}
where $\bar{E}({\mathbf{k}_{c}})=\text{Re}[E_\theta({\mathbf{k}_{c}})]+\bar{%
\mathbf{A}}_\theta(\mathbf{k}_{c}) \cdot {\dot {\mathbf{k}}}_{c}$, $\bar{%
\mathbf{A}}_\theta(\mathbf{k}_{c})\equiv\text{Re}[\mathbf{A}_\theta(\mathbf{k%
}_{c})-\tilde{\mathbf{A}}_\theta(\mathbf{k}_{c})]$ with the Berry connection
defined as $\mathbf{A}_\theta(\mathbf{k})=i\langle{u}_\theta(\mathbf{k}%
)|\partial_{\mathbf{k}}u_\theta(\mathbf{k})\rangle$ and $\tilde{\mathbf{A}}%
_\theta(\mathbf{k})=i\langle\tilde{u}_\theta(\mathbf{k})|\partial_{\mathbf{k}%
}u_\theta(\mathbf{k})\rangle$.
Here we suppose that the dynamics of a system is determined by the right state.
It is easily seen that the Berry curvature contributes
a transverse anomalous velocity, which plays the same role as in the conventional semiclassical
equation in a closed system~\cite{Niu1999}. Different from the closed system,
the gradient field modifies the energy spectrum due to the difference
between left and right eigenstates in a non-Hermitian system. Inside the Weyl
exceptional ring in the $k_z=0$ plane, the real part of dispersion vanishes, leading to
a zero group velocity of a wave packet if $\mathbf{F}$ is absent.

Besides the Chern number on a closed surface, we can characterize the Weyl exceptional ring
by the Berry phase, defined by
\begin{equation}
C_1=\oint_{2\mathcal{L}}{i\langle\tilde{u}_\theta(\mathbf{k})|\partial_{%
\mathbf{k}} u_\theta(\mathbf{k})\rangle} \cdot d{\bm k}.  \label{BerryEq}
\end{equation}
Compared to the Berry phase defined in the nodal ring system, 
where a state can return by travelling across the ring once~\cite{Burkov2011PRB,Yong2016PRATypeII,DWZhang2016},
in the Weyl exceptional ring, one needs the trajectory $2\mathcal{L}$ to travel
across the ring twice along the Riemann surface so that the state ends up
with the original state after the entire path [see Fig.~\ref{WER}(c) and (d)].
This definition gives the quantized Berry phase $C_1=\pm\pi$, which is in agreement
with the result for a single exceptional point~\cite{Moiseyev2011}.

The Weyl exceptional ring possesses both a nonzero quantized Chern number and a nonzero quantized Berry phase.
This is in stark contrast to the Weyl nodal line, which exhibits a quantized
Berry phase but a zero Chern number. When the non-Hermitian term is included in the multiple Weyl point
Hamiltonian (\ref{MWHam}), a more complex structure was found~\cite{Shanhui2018PRB}.
To realize a Weyl exceptional ring, there are proposals with cold atoms~\cite{Xu2017PRL} and with
photonic systems~\cite{Shanhui2018PRB}.
It was also shown that such a exceptional ring may exist in a disordered type-II
Weyl semimetal~\cite{Zyuzin2018PRB}. Recently, a Weyl exceptional ring has
been engineered and observed in an optical waveguide array~\cite{Rechtsman2018}.

\section{Gapless points and rings in superfluids}
In 1987, Volovik predicted that Weyl points can exist in quasiparticle spectra of He$^3$ A superfluid
phase. Compared with the strongly correlated system, cold atom systems
provide a flexible and controllable platform with simple and weak interactions. In this section,
we review the Weyl points, structured Weyl points, nodal rings, and structured nodal rings in quasiparticle spectra of
spin-orbit coupled superfluids and dipolar superfluids in cold atoms.

\subsection{Weyl points and structured Weyl points}
\subsubsection{Spin-orbit coupled superfluids}
In cold atoms, both 1D and 2D spin-orbit coupling has been experimentally
realized using Raman laser beams (see review for details~\cite{Spielman2013Nature,HuiZhai2015,XJLiu2018Review}). Let us first
consider the Fermi gases with 2D Rashba-type spin-orbit coupling and
an attractive $s$-wave contact interaction, where the Weyl points were found~\cite{Gong2011prl}.
The system is
described by the following many-body Hamiltonian
\begin{eqnarray}
H_S=&&\int d\mathbf{r}\hat{\Psi}^{\dagger }(\mathbf{r})H_{s}(\hat{%
\mathbf{p}})\hat{\Psi}(\mathbf{r})- \nonumber \\
&&U\int d\mathbf{r}\hat{\Psi}_{\uparrow
}^{\dagger }(\mathbf{r})\hat{\Psi}_{\downarrow }^{\dagger }(\mathbf{r})\hat{%
\Psi}_{\downarrow }(\mathbf{r})\hat{\Psi}_{\uparrow }(\mathbf{r}).
\end{eqnarray}
Here $H_s(\hat{\bf p})=\varepsilon _{\hat{\bf p}}-\mu+\alpha ({\bm\sigma }\times
{\hat{\bf p}})\cdot \hat{z}/\hbar$ is the single particle Hamiltonian with $\varepsilon_{\hat{\bf p}}=\frac{\hat{\bf p}^2}{2m}$, $m$ being the atom mass,
$\mu$ being the chemical potential and $\alpha$ characterizing the
spin-orbit coupling strength; $U$ describes the strength of attractive interactions;
$\hat{\Psi}(\mathbf{r}%
)=[\hat{\Psi}_{\uparrow }(\mathbf{r}),\hat{\Psi}_{\downarrow }(\mathbf{r}%
)]^{T}$ with $\hat{\Psi}_{\nu }^{\dagger }(\mathbf{r})$ [$\hat{\Psi}_{\nu }(%
\mathbf{r})$] being the fermionic atom creation (annihilation) operator.
Using the mean field approximation, we obtain
the
$4\times 4$ Bogoliubov-de
Gennes (BdG) Hamiltonian
\begin{eqnarray}
H_{BdG1} =H_s(\hat{{\bf p}}\rightarrow{\bf k})\tau _{z}+\Delta _{0}\tau _{x} +h_{z}\sigma _{z},
\label{BdG}
\end{eqnarray}
where $\sigma_\nu$ and $\tau_\nu$ ($\nu=x,y,z$) are the Pauli matrices,
acting on the spin and Nambu spaces, respectively,
$\Delta_0=-U\langle \hat{%
\Psi}_{\downarrow }(\mathbf{r})\hat{\Psi}_{\uparrow }(\mathbf{r}) \rangle$ is the order parameter
arising from the pairing instability because of attractive interactions, and $h_z$
denotes the strength of the out-of-plane Zeeman field. The system possesses an intrinsic
particle-hole symmetry, i.e., $\Xi^{-1}H_{BdG1}({\bf k})\Xi=-H_{BdG1}(-{\bf k})$ with
$\Xi =\sigma _{y}\tau _{y}K$ and $K$ being the complex conjugation ($\Xi^2=1$), implying that
if there exist an eigenstate $|u_n({\bf k})\rangle$
with energy being $E_n(\bf k)$, there exist another eigenstate $\Xi|u_n({\bf k})\rangle$ whose
energy is $-E_n(\bf k)$.

The quasiparticle spectrum is given by
\begin{equation}
E_{\pm}^{\lambda}({\bf k})=\lambda\sqrt{ \Lambda_{\bf k}^2 +\alpha^2k_\perp^2+h_z^2 \pm
2\sqrt{h_z^2\Lambda_{\bf k}^2+\alpha^2 k_\perp^2\xi_{\bf k}^2}},
\end{equation}
with $\Lambda_{\bf k}^2=\xi_{\bf k}^2+|\Delta_0|^2$, $\xi_{\bf k}=\varepsilon_{\bf k}-\mu$,
$k_\perp=\sqrt{k_x^2+k_y^2}$, and
$\lambda=\pm$ indicating the particle ($+$) and hole branches ($-$) respectively.
Clearly, without the spin-orbit coupling and the Zeeman
field ($\alpha=h_z=0$), $E_{{\bf k},\pm}^\lambda=\lambda |\Lambda_{\bf k}|$,
which is the quasiparticle spectrum of a typical BCS superfluid.
To see whether a gap closing point appears, let us compute the product of
the two branches of energies, which gives
\begin{equation}
(E_{+}^{+}({\bf k})E_{-}^{+}({\bf k}))^2
=\left(h_z^2+\alpha^2k_\perp^2-\Lambda_{\bf k}^2\right)^2+4\alpha^2k_\perp^2|\Delta_0|^2.
\end{equation}
Evidently, when $k_x=k_y=0$ and $\xi_{\bf k}^2+|\Delta_0|^2=h_z^2$, the gap closes.
This occurs when $h_z\ge |\Delta_0|$ for $\mu>0$. If $|\Delta_0|<h_z<\sqrt{\mu^2+|\Delta_0|^2}$,
there are four gapless points at $k_{z}=k_c=\pm\sqrt{\frac{2m}{\hbar^2}(\mu+\sqrt{h_z^2-|\Delta_0|^2})}$ and
$k_{z}=k_c=\pm\sqrt{\frac{2m}{\hbar^2}(\mu-\sqrt{h_z^2-|\Delta_0|^2})}$; if $h_z>\sqrt{\mu^2+|\Delta_0|^2}$,
there are two points at $k_{z}=k_c=\pm\sqrt{\frac{2m}{\hbar^2}(\mu+\sqrt{h_z^2-|\Delta_0|^2})}$~\cite{Gong2011prl,YongReveiw}.

Near the zero energy point ${\bf k}={\bf k}_W$ with ${\bf k}_W=(0,0,k_{c})$,
the energy can be expanded with respect to $\delta k_x$, $\delta k_y$ and $\delta k_z$ to the first order,
\begin{equation}
E_{-}^{\pm}=\pm \left( v_{x}|\delta k_x|+v_{y}|\delta k_y|+
v_z|\delta k_z| \right)+O(\delta{\bf k}^2),
\end{equation}
where $\delta k_x$, $\delta k_y$ and $\delta k_z$ are measured with respect to ${\bf k}={\bf k}_W$, and
$v_{x}=v_y=\frac{\alpha\Delta_0}{h_z}$ and $v_{z}=\frac{\hbar^2|\xi_{{\bf k}_W}k_{c}|}
{m\sqrt{\Delta_0^2+\xi_{{\bf k}_W}^2}}$. The spectrum is linear along all three directions,
indicating that these gapless points are Weyl points. The anisotropy of
the parameters $v_x=v_y$ and $v_z$ appears because the spin-orbit coupling only exist in
the $(x,y)$ plane. We note that the linear part of the dispersion along the z direction vanishes
at the transition point where $k_c=0$ or $\xi_{{\bf K}_W}=0$, suggesting the absence of Dirac
points that feature the linear dispersion.

In the presence of an in-plane Zeeman field $h_x\sigma_x$, the Cooper pairs pick up a finite
center-of-mass momentum~\cite{Zheng2013PRA,FanWu2013PRL} since the Zeeman field makes the Fermi surface asymmetric along
the $y$ direction~\cite{YongPRA2013}, i.e., the order parameter takes the form of
$\Delta({\bf r})=\Delta_0 e^{i Q_y y}$ with $\hbar Q_y$ denoting the finite momentum of a
Cooper pair (such superfluids with a finite center-of-mass momentum of Cooper pairs is called
Fulde-Ferrell superfluids~\cite{FuldePR1964,Larkin1964}). In this case, the BdG
Hamiltonian becomes
\begin{eqnarray}
H_{BdG2} &=&\left[ \varepsilon _{\bm k}-\bar{\mu}+\alpha ({\bm\sigma }\times
{\bm k})\cdot \hat{z}\right] \tau _{z}+\Delta _{0}\tau _{x}  \notag \\
&+&\bar{h}_{x}\sigma _{x}+h_{z}\sigma _{z}+{\hbar ^{2}k_{y}Q_{y}}/{2m},
\label{SOCFFBdG}
\end{eqnarray}
with $\bar{\mu}=\mu -Q_{y}^{2}\hbar ^{2}/8m$ and $\bar{h}_{x}=h_{x}+\alpha
Q_{y}/2$.

\begin{figure}[t]
\includegraphics[width=3.2in]{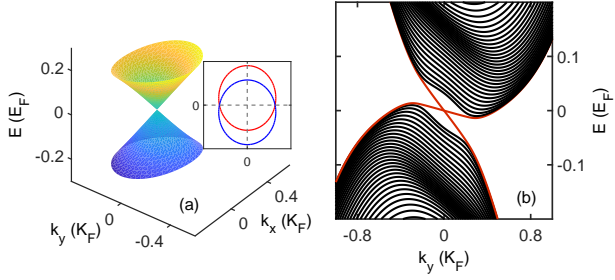}
\caption{(Color online) (a) Quasiparticle energy spectra near a Weyl point for a fixed $k_z$.
The inset plots the energy contours for $E=0.1E_{F}$ (red line) and $E=-0.1E_{F}$ (blue line).
(b) Quasiparticle energy spectra
in the gapless topological Fulde-Ferrell phase with respect to $k_{y}$ for a
fixed $k_z$ under an open boundary in the $x$ direction. Black and red lines
correspond to the bulk and surface states, respectively. Here, $E_{F}$ and $K_F$ refer to
the Fermi energy and Fermi vector, respectively.
Reproduced from Ref. [\cite{Xu2014PRL}].}
\label{gaplessWeyl}
\end{figure}

Let us analyze the symmetry of the system.
It possesses a combined symmetry $\Pi_y =\Xi \Theta \mathcal{M}_{y}=i\sigma _{y}\tau _{y}$, i.e.,
$\Pi_y ^{-1}H_{BdG2}\Pi_y
=-H_{BdG2}(-k_{y})$, where $\Theta =\sigma_{y}K$ denotes the time-reversal operator,
and $\mathcal{M}_{\nu }=-i\sigma
_{\nu }$,\thinspace $\nu =x,y$ denote the mirror symmetries, both of which are broken
in the presence of Zeeman fields. At $k_y=0$, $\Pi_y ^{-1}H_{BdG2}\Pi_y
=-H_{BdG2}$, indicating that the gapless points in this plane, if exist, have to be
doubly degenerate. Numerical calculation further shows that the degenerate gapless points occur
on the line $k_x=k_y=0$. As a result, Weyl points appear when
\begin{equation}
\bar{h}_x^{2}+h_{z}^{2}=(\hbar ^{2}k_{z}^{2}/2m-\bar\mu
)^{2}+\Delta _{0}^{2}.
\label{zero_pts}
\end{equation}%
From this equation, we obtain the location $\mathbf{k}_{W}$
of Weyl points.
Similar to the system with pure $h_z$,
when $h_{z}^2>
\bar\mu ^{2}+\Delta _{0}^{2}-\bar{h}_x^{2}$, there are two gapless points
and when $\Delta _{0}^{2}-\bar{h}_x^{2}<h_{z}^2<\bar\mu%
^{2}+\Delta _{0}^{2}-\bar{h}_x^{2}$ and $\bar\mu>0$, there are
four zero excitations. The presence of $h_x$ decreases both critical values for
$h_z$ and when $h_x=\pm\Delta_0-\alpha Q_y/2$, there are two gapless points at $h_z=0$.
This shows that the location and number of Weyl points can be controlled by tuning the
Zeeman fields. The linear dispersion along the $k_x$ and $k_y$ is shown in Fig.~\ref{gaplessWeyl}
and it is anisotropic along all
three dimensions due to the presence of $h_x$.

\begin{figure}[t]
\includegraphics[width=3.2in]{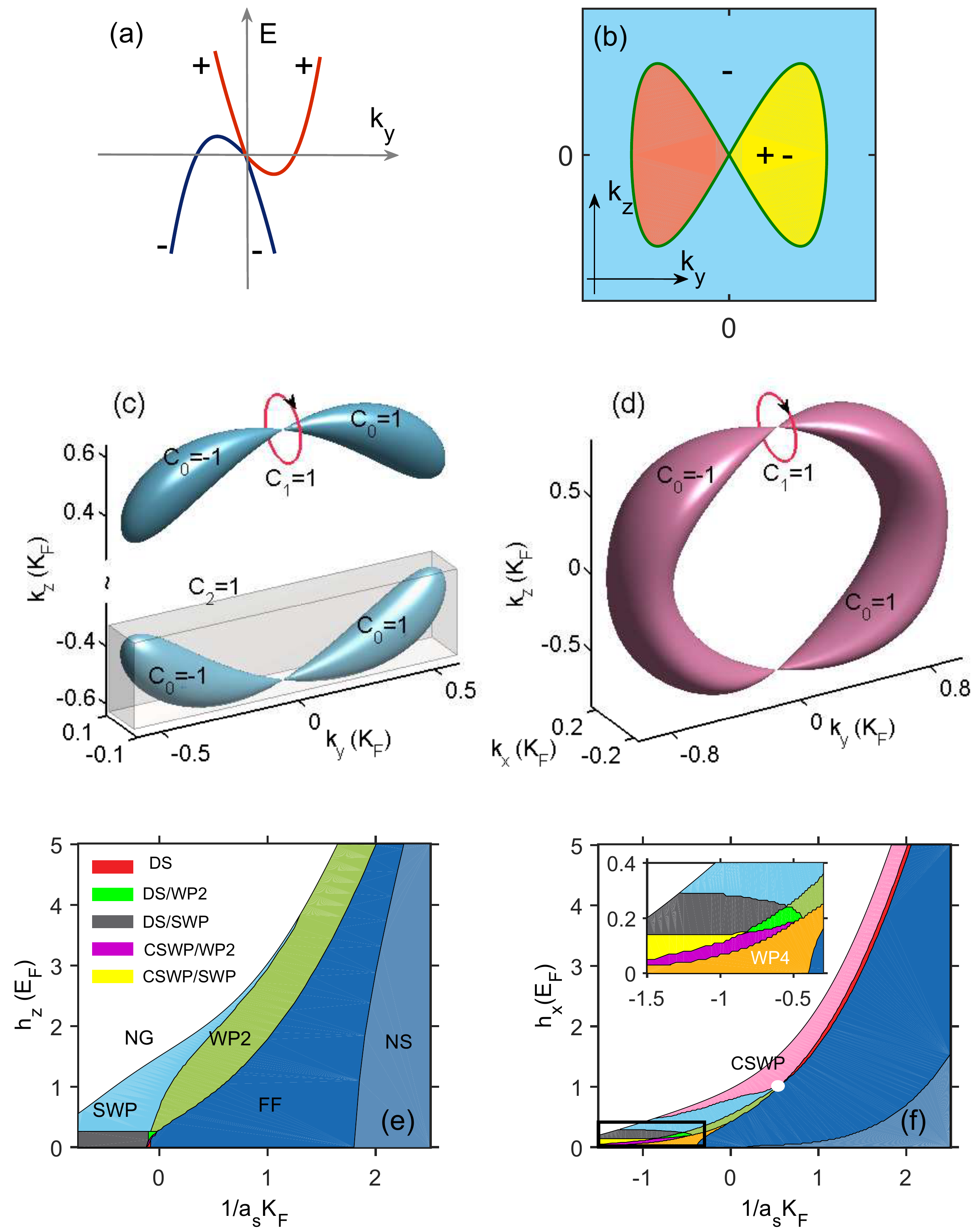}
\caption{(Color online) (a) The energy spectrum along $k_y$ of a structured Weyl point described by the Hamiltonian
(\ref{SWHSuperfluid}) for $k_x=k_z=0$. (b) Profiles of band occupation in the $k_x=0$ plane.
Fermi surfaces (the zero-energy contours) for a pair of structured Weyl points
in (c) and for a connected structured Weyl point in (d). Phase diagrams across the BCS-BEC crossover for a fixed $h_x$ in
(e) and for a fixed $h_z$ in (f). The inset is the zoomed-in view of the black rectangular area.
WP2: two Weyl points; WP4: four Weyl points; SWP: structured
Weyl point; CSWP: connected structured Weyl points; DS: disconnected
spheres; NS: normal superfluids; NG: normal gases.
The phase name with two parts is used to show the gapless structure for the inner and
outer touching points, respectively, for the superfluids with four touching points.
Adapted from Ref. [\cite{Xu2015PRL}].
}
\label{StructuredPD}
\end{figure}

In the absence of $h_x$ (thus $Q_{y}=0$), the system has an extra symmetry $%
\mathcal{M}=\Theta \mathcal{M}_{x}$, $%
\mathcal{M}^{-1}H_{BdG2}\mathcal{M}=H_{BdG2}(-k_{y},\pm k_{z})$ in light of
the intrinsic symmetry $H_{BdG2}=H_{BdG2}(-k_{z})$. We have
$({\Pi}_y \mathcal{M})^{-1}H_{BdG2}\Pi_y \mathcal{M}=-H_{BdG2}(k_{y},\pm k_{z})$
(similar to the chiral symmetry), which guarantees the double degeneracy for the
zero energy states if exist.
However,
with $h_x$, the $\mathcal{M}$ symmetry is broken, and the non-degenerate gapless points can appear
when $k_y\neq 0$ (in the $k_y=0$ plane, the $\Pi_y$ symmetry can still ensure the double degeneracy).
This makes it possible to realize the structured
Weyl point, where there exist gapless points which are not degenerate. Indeed, when $h_x$ is sufficiently large,
a Weyl point develops into a structured Weyl point~\cite{Xu2015PRL} characterized by
\begin{equation}
H_{SWP}=-\alpha k_{y}\sigma _{0}+(k_{y}+\gamma k_{y}^{3})\sigma
_{y}+k_{x}\sigma _{x}+k_{z}\sigma _{z}.  \label{SWHSuperfluid}
\end{equation}%
Compared to the toy model in (\ref{SWToy}), there appears a higher order term $\gamma k_y^3$.
Its appearance is due to the finite region in momentum space that the spin-orbit coupling takes effect in.
This means that the quasiparticle spectrum for large $k_\perp$ in the continuous system without lattices
is almost identical to that without spin-orbit coupling with two hole bands being occupied, which
is impossible without including the nonlinear term. With this term, the Fermi surface of the
toy model is consistent with that of the quasiparticle spectrum of the Fulde-Ferrell superfluids [see Fig.~\ref{StructuredPD}].
Fig.~\ref{StructuredPD} also illustrated a very rich phase diagram of the superfluids.

\begin{figure}[t]
\includegraphics[width=3.2in]{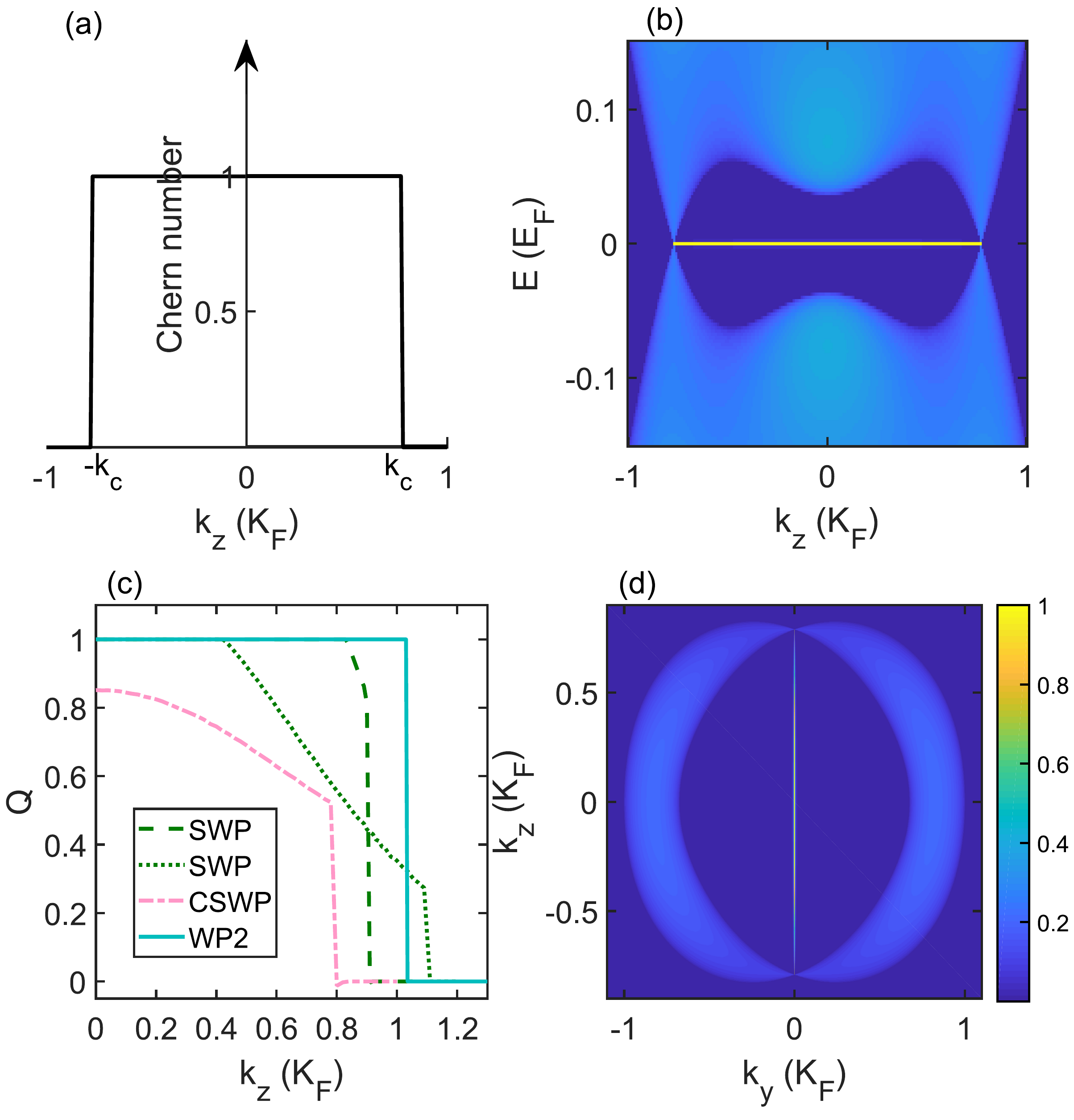}
\caption{(Color online) (a) The Chern number in the $(k_x,k_y)$ plane for a fixed $k_{z}$.
(b) The density of states with respect to $k_z$ and energy $E$ for $k_{y}=0$ with a
confinement along the $x$ direction. The yellow line shows the Fermi arc while
the light blue region represents the bulk states.
(c) Integral of Berry curvatures for the occupied bands in the $(k_x,k_y)$ plane
for a fixed $k_z$. Viewing $k_z$ as a parameter, such an integral shows the thermal
Hall effect in 2D superfluids and  the anomalous Hall effects in 2D non-interacting systems.
(d) The density of states in the $(k_y,k_z)$ plane for $k_x=0$ at zero energy for a connected
structured Weyl point, two connected structured Weyl points.
(a-b) are reproduced from Ref. [\cite{Xu2014PRL}];
(c-d) are adapted from the arXiv version of Ref. [\cite{Xu2015PRL}];
}
\label{SOCFermiarc}
\end{figure}

The structured Weyl point can still be characterized by the first Chern number defined
over a closed surface with finite energy gap enclosing the structure. However,
it is possible that such a gapped surface does not exist [see Fig.~\ref{StructuredPD}(d)
for example]. In this case, we can define the Chern number over the hole
band regardless of whether it is occupied or not. Actually, we can
always define the Chern number following this method and see that the Fermi arc
appears connecting two doubly degenerate point of structured Weyl points~\cite{Xu2015PRL,Soluyanov2015Nature}
[see Fig.~\ref{SOCFermiarc}].

As discussed in the preceding section, a Weyl point can still be regarded as the topological phase transition point
between topologically nontrivial and trivial insulators in momentum space.
In the spin-orbit coupled superfluid, the Hamiltonian (\ref{SOCFFBdG})
can be regarded as stacking of 2D layers in terms of the parameter $k_z$.
Each layer corresponds to a Chern insulator with the Chern number defined by
\begin{equation}
C\left( k_{z}\right) =\frac{1}{2\pi }\sum_{n}\int dk_{x}dk_{y}\Omega
^{n}(k_{x},k_{y}),  \label{Chern_equ}
\end{equation}%
where $n$ is the index for hole branches, and the Berry curvature in the
z direction can be written as \cite{XiaoRMP}
\begin{equation}
\Omega ^{n}=i\sum_{n^{\prime }\neq n}\left[ \frac{\langle n|\partial
_{k_{x}}H|n^{\prime }\rangle \langle n^{\prime }|\partial
_{k_{y}}H|n\rangle -(k_{x}\leftrightarrow k_{y})}{(E_{n\mathbf{k}%
}-E_{n^{\prime }\mathbf{k}})^{2}}\right]
\end{equation}%
with $n^{\prime }$, which is not equal to $n$, running over the eigenstates of $%
H$. As shown in Fig.~\ref{SOCFermiarc}(a),
when $-k_c<k_z<k_c$ (two Weyl points are located at $\pm k_c$), $C=1$, implying a topologically nontrivial phase, and otherwise,
$C=0$, implying a topologically trivial one. The presence of the Chern number gives rise to the chiral Majorana edge bands, which leads to
the flat Majorana arc at zero energy connecting two Weyl points as shown in
Fig.~\ref{SOCFermiarc}(b).

When the Weyl point develops into a structured Weyl point,
the definition in Eq.(~\ref{Chern_equ}) yields a crossover of the Chern number
from 0 to 1, taking nonquantized values. The Chern number is not quantized because only
the contribution from the occupied bands is integrated, i.e.,
the integral surface is not closed due to the gapless structure of
a structured Weyl point [see Fig.~\ref{StructuredPD}(b)].
In 2D, this effect can manifest as
the anomalous Hall effect in a non-interacting system and as
the thermal Hall effect in superfluids. In an open geometry, chiral edge states
appear. Remarkably, the group velocities are along the same direction on the opposite surfaces, in
contrast to the conventional case with opposite direction [see Fig.~\ref{gaplessWeyl}(b)].
Such a topological gapless phenomenon was further studied in the 2D
spin-orbit coupled Fulde-Ferrell superfluids~\cite{Hu2014PRL,Xu2015PRLBKT,HuiHu2015}. Recently, this 2D
topological gapless phenomenon (termed topological metal) was also found in a Fulde-Ferrell $p+ip$ superconductor
and the nonquantized thermal Hall effects were predicted~\cite{Kamenev2018PRL}.

As we have mentioned, the edge states emerge even though the system becomes gapless.
In this case, we can define the Chern number by the hole bands instead of using
the occupied bands. From this definition, we have quantized Chern numbers for the tilted
hole and particle bands and chiral edge mode appears inside their gap. These surface
states lead to the Fermi arc (Majorana zero modes in superfluids) connecting two degenerate
points even though they have
developed into two structured Weyl points connected together [see Fig.~\ref{SOCFermiarc}(d)].

Such gapless topological Fulde-Ferrell superfluids with structured
Weyl fermions can also exist in the attractive Fermi gases with the Weyl spin-orbit
coupling~\cite{Huihu2014PRA}. Since there is no distinction between in-plane and out-of-plane
Zeeman fields, the gapless topological Fulde-Ferrell phase appears with only one
Zeeman field. Furthermore, spiral Majorana zero modes forwarding on the surfaces were found
in optical lattices with the Weyl spin-orbit coupling, where the Weyl points
are located at different energies~\cite{KTLaw2018PRA}.

\subsubsection{dipolar superfluids}
In addition to the spin-orbit coupled superfluids, a 3D single-component dipolar
Fermi gas was proposed to exhibit Weyl points in the quasiparticle spectrum~\cite{Liubo2015PRL}.
Such a superfluid may be realized in magnetic dipolar atoms such as $^{167}$Er and $^{161}$Dy
or polar molecules. By applying an external fast rotating magnetic field, all dipoles are aligned along the
magnetic field direction with the effective dipole-dipole interactions taking the following form
\begin{equation}
V({\bf r})=\frac{d^2(3\cos^2\phi-1)}{2r^3}(1-3\cos^2\theta),
\end{equation}
with the magnetic dipole moment $d$, the distance $r$ between two dipoles,
the relative angle $\theta$ between $\bf r$ and the $z$ direction, $\phi$ is
the angle between the magnetic field and the $z$ direction. When $\phi$ is
controlled such that $V({\bf r})<0$, the interaction becomes attractive and
the pairing instability occurs.

This system can be described by the following many-body Hamiltonian,
\begin{eqnarray}
H_{DD}&&=\int d\mathbf{r}\hat{\Psi}^{\dagger }(\mathbf{r})(-\frac{\hbar^2\nabla^2}{2m}-\mu )\hat{\Psi}(\mathbf{r})- \nonumber \\
&&\frac{1}{2}\int d\mathbf{r}\int d{\bf r}^\prime \hat{\Psi}^{\dagger }(\mathbf{r})\hat{\Psi}^{\dagger }(\mathbf{r}^\prime)
V({\bf r}-{\bf r}^\prime)\hat{%
\Psi}(\mathbf{r}^\prime)\hat{\Psi}(\mathbf{r}).
\end{eqnarray}
Let us define the mean-field order parameter $\Delta({\bf r}-{\bf r}^\prime)=\langle V({\bf r}-{\bf r}^\prime)\hat{%
\Psi}(\mathbf{r}^\prime)\hat{\Psi}(\mathbf{r})\rangle$. Since the dominant contribution of the pairing comes
from the pairing with orbital angular momentum being $1$, let us consider the $p_x+ip_y$ pairing (which is
more stable than $p_x$ or $p_y$ phase~\cite{Liubo2015PRL}) and the order
parameter in momentum space takes the form $\Delta_{\bf k}=\Delta(k_\rho,k_z)e^{i\phi_{\bf k}}$ with
$k_\rho=\sqrt{k_x^2+k_y^2}$ and $\phi_{\bf k}$ being the polar angle of the momentum $\bf k$ in the
$(k_x,k_y)$ plane. Here $\Delta_{k_\rho,k_z}=-\Delta_{-k_\rho,k_z}$, thus $\Delta_{k_\rho=0,k_z}=0$.
In the momentum space, the BdG Hamiltonian can be written as
\begin{equation}
H_{DD}({\bf k})=\frac{1}{2}\left(
    \begin{array}{cc}
      \xi({\bf k}) & \Delta({\bf k}) \\
      \Delta({\bf k})^* & -\xi({\bf k}) \\
    \end{array}
  \right),
\end{equation}
where $\xi({\bf k})=\frac{\hbar^2 k^2}{2m}-\mu$ if we do not consider the Hartree-Fock self-energy, which
will not change the physics qualitatively. At $k_\rho=0$, $\Delta=0$, a degenerate point appears at
$k_z=\pm\sqrt{2m\mu/\hbar^2}$. If we simply write $\Delta_{\bf k}=\Delta_0(k_x+i k_y)$, the linear dispersion
is easily seen. Hence, a Weyl point is achieved in the quasiparticle spectrum of a dipolar superfluid.

\subsection{Nodal rings}
\subsubsection{Spin-orbit coupled superfluids}
Seo and coworkers found that nodal rings can emerge in
quasiparticle spectra of Fermi superfluids with equal Rashba and
Dresshauls spin-orbit coupling~\cite{Melo2012PRL}. Considering
this type of spin-orbit coupling,
we have the following BdG Hamiltonian
\begin{eqnarray}
H_{BdG} &=&\left[ \varepsilon _{\bm k}-\bar{\mu}+\alpha k_y\sigma_x\right] \tau _{z}+\Delta _{0}\tau _{x}  \notag \\
&+&\bar{h}_{x}\sigma _{x}+h_{z}\sigma _{z}+{\hbar ^{2}k_{y}Q_{y}}/{2m},
\label{BdGRing}
\end{eqnarray}
Similar to the Hamiltonian (\ref{SOCFFBdG}) for the Rashba spin-orbit coupling, without $h_x$
($Q_y=0$ when $h_x=0$), the system has both the $\Pi_y$ and $\mathcal{M}$ symmetry so that
$(\Pi_y \mathcal{M})^{-1}H_{BdG}\Pi_y \mathcal{M}=-H_{BdG}$. This chiral symmetry ensures that
a gapless point, if exists, is at least doubly degenerate. Specifically, the quasiparticle spectrum is
\begin{equation}
E_{\pm}^{\lambda}({\bf k})=\lambda\sqrt{ \Lambda_{\bf k}^2 +\alpha^2k_y^2+h_z^2 \pm
2\sqrt{h_z^2\Lambda_{\bf k}^2+\alpha^2 k_y^2\xi_{\bf k}^2}},
\end{equation}
leading to
\begin{equation}
(E_{+}^{+}({\bf k})E_{-}^{+}({\bf k}))^2
=\left(h_z^2+\alpha^2k_y^2-\Lambda_{\bf k}^2\right)^2+4\alpha^2k_y^2|\Delta_0|^2.
\end{equation}
Apparently, the nodal rings emerge in the $k_y=0$ plane described by $\xi_{\bf k}^2+|\Delta_0|^2=h_z^2$
(i.e., $k_x^2+k_z^2=\frac{2m}{\hbar^2}(\mu \pm \sqrt{h_z^2-|\Delta_0|^2})$), showing that
there is one ring when $h_{z}^{2}>{\mu}%
^{2}+|\Delta _{0}|^{2}$, and two rings when $|\Delta
_{0}|^{2}<h_{z}^{2}<{\mu}^{2}+|\Delta _{0}|^{2}$ and ${\mu}>0$.

\begin{figure}[t]
\includegraphics[width=3.2in]{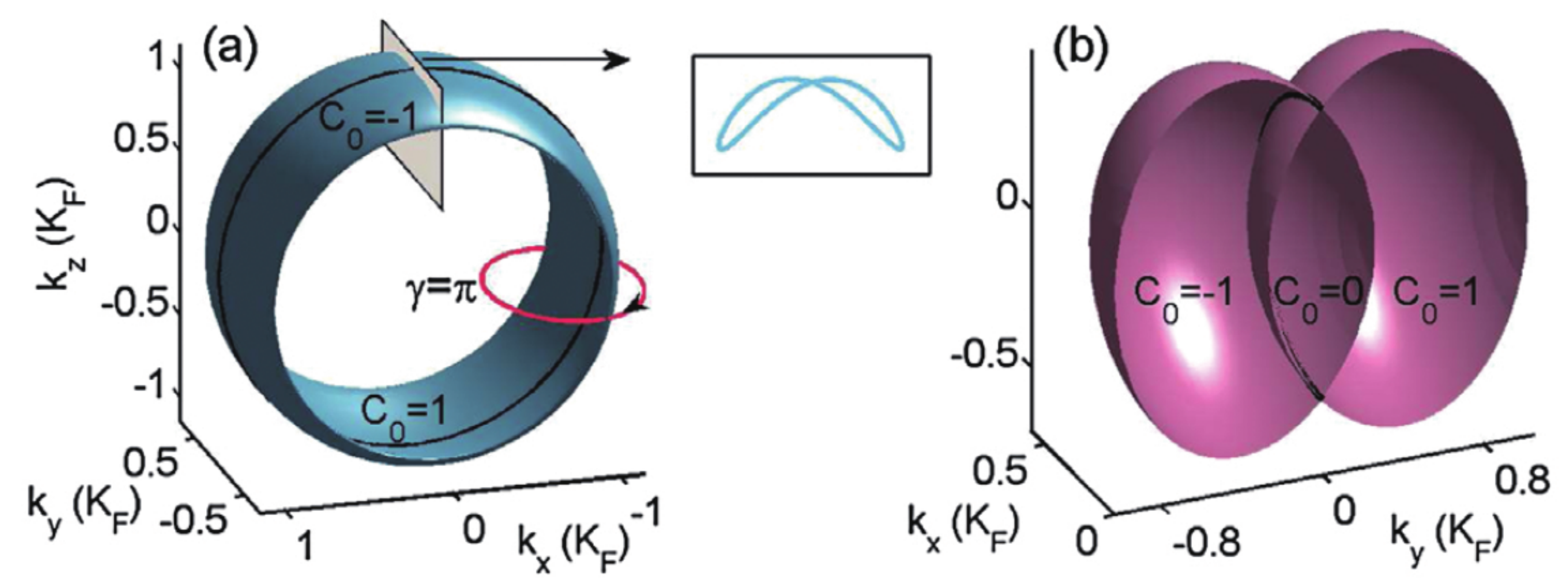}
\caption{(Color online)
(a) The Fermi surface (zero-energy contour) of a structured Weyl ring. The inset
plots the cross section of a structured Weyl ring. (b) The Fermi surface of a
closed structured Weyl ring for $k_x>0$.
Reproduced from Ref. [\cite{Xu2015PRL}].}
\label{StructuredNodalRing}
\end{figure}

With the in-plane Zeeman field $h_x\sigma_x$, the 3D superfluids become Fulde-Ferrell type with
nonzero $Q_y$~\cite{Xu2015PRL}. This term
breaks the $\mathcal{M}$ symmetry, so that the gapless point does not have to be
doubly degenerate. However, the system still respects the $\Pi_y$ symmetry, meaning that in the $k_y=0$
plane any zero energy states are still at least doubly degenerate.
Furthermore, because
of the rotational symmetry with respect to $%
k_{y}$ [$H_{BdG}=H_{BdG}(k_{x}^{2}+k_{z}^{2},k_{y})$], a degenerate ring can
appear in the $k_y=0$ plane. Additionally, it is possible for the non-degenerate gapless points
to appear away from the $k_y=0$ plane. Indeed, the structured Weyl ring was found, which
is described by the effective Hamiltonian~\cite{Xu2015PRL}
\begin{equation}
H_{SWR}=-\alpha
k_{y}\sigma _{0}+(k_{y}+\gamma k_{y}^{3})\sigma _{y}+(\mathbf{k}%
^{2}-m^{2})\sigma _{z}
\label{SWRSupEff}
\end{equation}
with nonzero $m$. See Fig.~\ref{StructuredNodalRing} for the gapless structure. The ring is protected by the
$PT$ symmetry [$\Theta_{PT}=\sigma_z\mathcal{K}$ for the model (\ref{SWRSupEff}) and $\Theta_{PT}=\mathcal{K}$
for the BdG Hamiltonian (\ref{BdGRing})] and the Berry phase for a hole band enclosing the ring
has to be 0 or $\pi$.

\begin{figure}[t]
\includegraphics[width=3.2in]{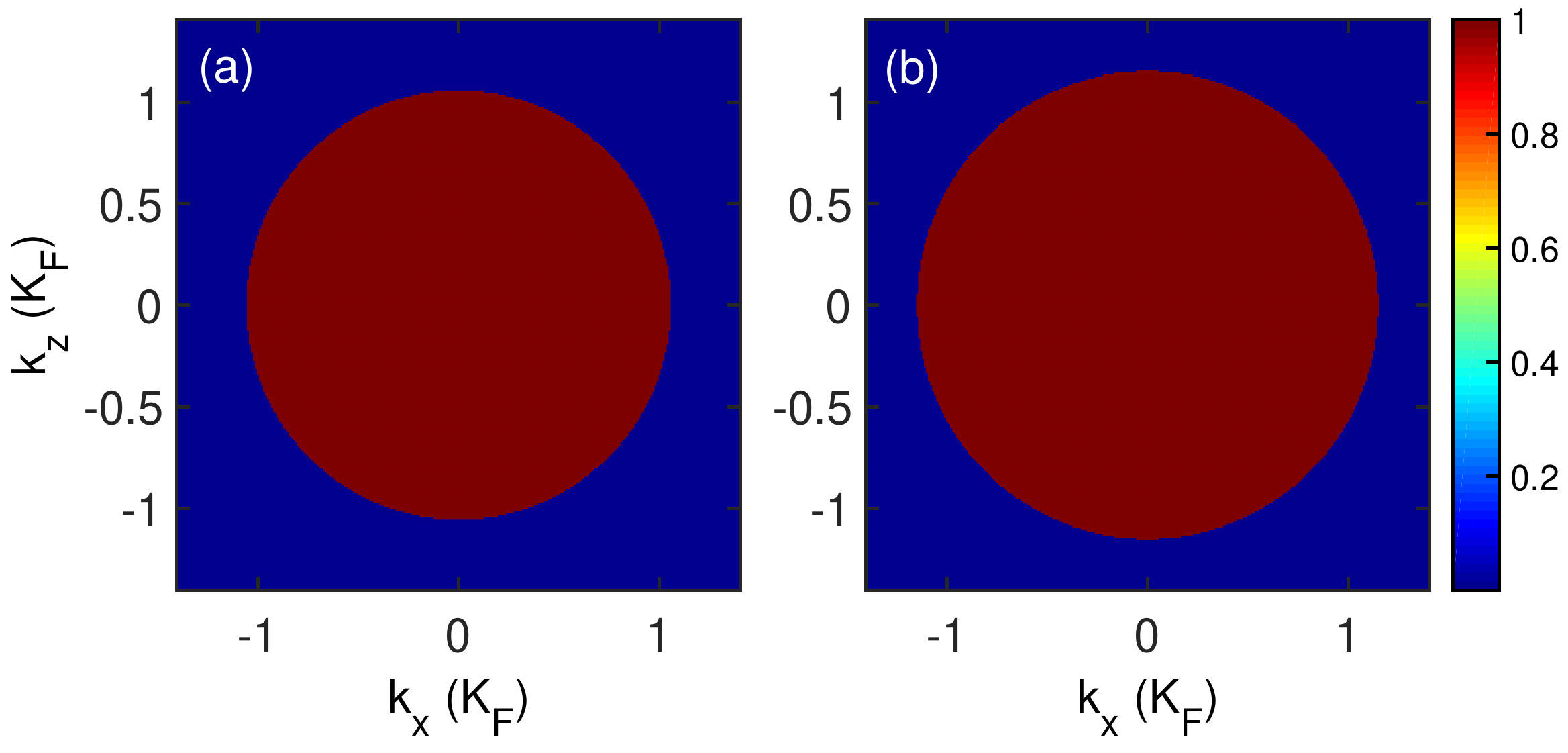}
\caption{(Color online) The density of states of the superfluids with a Weyl nodal ring in (a) and a structured
Weyl nodal ring in (b) under an open boundary along the $y$ direction.
Adapted from the arXiv version of Ref. [\cite{Xu2015PRL}].}
\label{FermiArc1DSoca}
\end{figure}

In an open geometry, Majorana flat bands occur not only in the case with $h_x=0$ for
traditional nodal rings, but also in the case with $h_x\neq 0$ for structured
nodal rings (see Fig.~\ref{FermiArc1DSoca}), as the topology is fully determined
by the hole band irrelevant of its occupation. Moreover, more than two nodal rings
were found in the quasiparticle spectrum of Fermi superfluids, when attractive
atoms were loaded into an optical lattice with Weyl nodal rings in the single
particle spectrum~\cite{Yong2016DiracRing}

\section{Gapless points and rings in non-interacting ultracold atomic gases}
Compared with a superfluid, a non-interacting system is easier to control
and probe in cold atom systems. Hence, similar to searching for specific solid-state materials
with topological gapless band structures, in cold atoms, there has been
considerable interest in looking for experimental schemes to realize these topological
gapless bands in non-interacting cold atom systems. Since the interaction can
be readily tuned by Feshbach resonances in cold atoms~\cite{Chin2010RMP},
non-interacting systems can be achieved there. Further, the effects of interactions
can be studied in the gapless system, when interactions are turned on.
In this section, we summarize a number of experimental proposals in cold atoms
for realization of Weyl points, structured (type-II) Weyl points,
nodal rings and Weyl exceptional rings.

\subsection{Weyl points}
\subsubsection{Weyl spin-orbit coupling}
Spin-orbit coupling, the interaction between spin and orbital degree of freedom, plays a key role
in many topological phenomena, such as topological insulators~\cite{Hasan2010RMP,ZhangRMP2011}, anomalous and
spin Hall effect~\cite{XiaoRMP},
and Majorana zero modes~\cite{Alicea2018}, \textit{etc}. In solid-state materials, electrons naturally experience the
spin-orbit coupling as an electron moves in an internal electric field arising from the atomic potential and
the crystal field. For cold atoms, they are neutral and do not experience any electric and
magnetic fields. However, since atoms can interact with laser beams, one can engineer these gauge fields
including the spin-orbit coupling using laser beams. To date, both 1D and 2D spin-orbit coupling has been engineered in cold atoms
~\cite{Spielman2011Nature,JYZhang2012PRL,Jing2012PRL,Zwierlein2012PRL,CQu2013PRA,Olson2014PRA,Luo2014SR,Lev2016PRX,Jo2016PRA,
Jo2016PRA,JLi2016PRL,Ye2017nature,JRLi2017Nature,Song2018SciAdv,Jing2016NP,Jing2016PRL,Xiongjun2015Sci,Xiongjun2018PRL,Song2018}.

Besides the 1D and 2D spin-orbit coupling, an experimental proposal to implement the 3D Weyl spin-orbit coupling was reported~\cite{Spielman2012PRL}, which is described by the effective Hamiltonian
\begin{equation}
H=\frac{{\bm p}^2}{2m}+\lambda {\bm p}\cdot{\bm \sigma}
\label{SpielmanHam}
\end{equation}
with $\lambda$ being a real parameter and $m$ being the mass of atoms.
Since ${\bm \sigma}$ are defined in the hyperfine levels, $\lambda {\bm p}\cdot{\bm \sigma}$ refers to
the 3D Weyl spin-orbit coupling. This term certainly describes a Weyl point.
Notice that there is only one single Weyl point, which is possible as the system is in a free
space without any lattice.

\begin{figure}[t]
\includegraphics[width=3.2in]{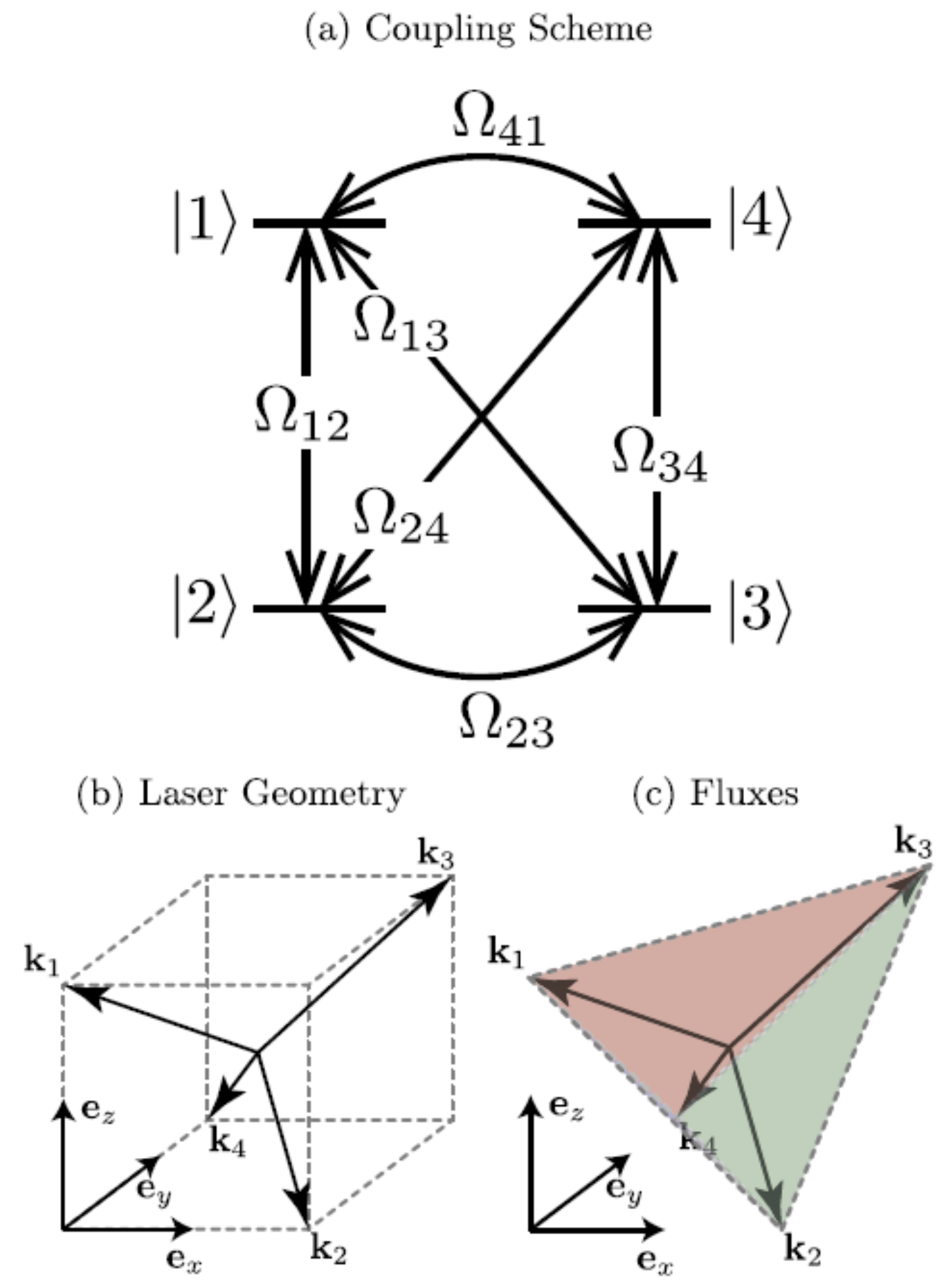}
\caption{(Color online) (a) The laser configuration for the 3D Weyl spin-orbit coupling, where
four levels are coupled using Raman laser beams. (b) The wave vector configuration of laser
beams. (c) The phase is is chosen as $\Phi_i=\sum_{k\neq i}\phi_{k,k+1}=\pi/2 \text{mod} 2\pi$.
Reproduced from Ref. [\cite{Spielman2012PRL}].}
\label{3DSOCSpielman}
\end{figure}

To realize the 3D Weyl spin-orbit coupling, a 4-level atom with states $|1\rangle$, $|2\rangle$, $|3\rangle$
and $|4\rangle$ coupled by Raman laser beams is considered; the system is described by the Hamiltonian
\begin{equation}
H_{al}=\sum_{jk}\Omega_{jk}|j\rangle\langle k|,
\end{equation}
where $\Omega_{jk}=\Omega^{(jk)}\text{exp}[i({\bf k}_{jk}\cdot {\bf r}+\phi_{jk})]$
with the transferred momentum ${\bf k}_{jk}={\bf k}_j-{\bf k}_k$ and ${\bf k}_j$ being the wave vector of laser beams,
and $\phi_{ik}$ is the coupling phase. In atom $^{87}$Rb, these levels can be chosen as: $|1\rangle=|F=2,m_F=0\rangle$, $|2\rangle=|F=1,m_F=1\rangle$,
$|3\rangle=|F=1,m_F=0\rangle$ and $|4\rangle=|F=2,m_F=1\rangle$. The coupling strength $\Omega^{(jk)}$
is chosen such that $\Omega^{(jk)}=\Omega^{(1)}$ for $k=j+1$ and $\Omega^{(jk)}=\Omega^{(2)}$ for $k=j+2$.
The momentum vectors are taken as
\begin{equation}
{\bf k}_j=\kappa_\perp ({\bf e}_x\cos\beta_j-{\bf e}_y\sin\beta_j)-\kappa_{||}(-1)^j{\bf e}_z,
\end{equation}
as shown in Fig.~\ref{3DSOCSpielman}(b).

\begin{figure*}[t]
\includegraphics[width=\textwidth]{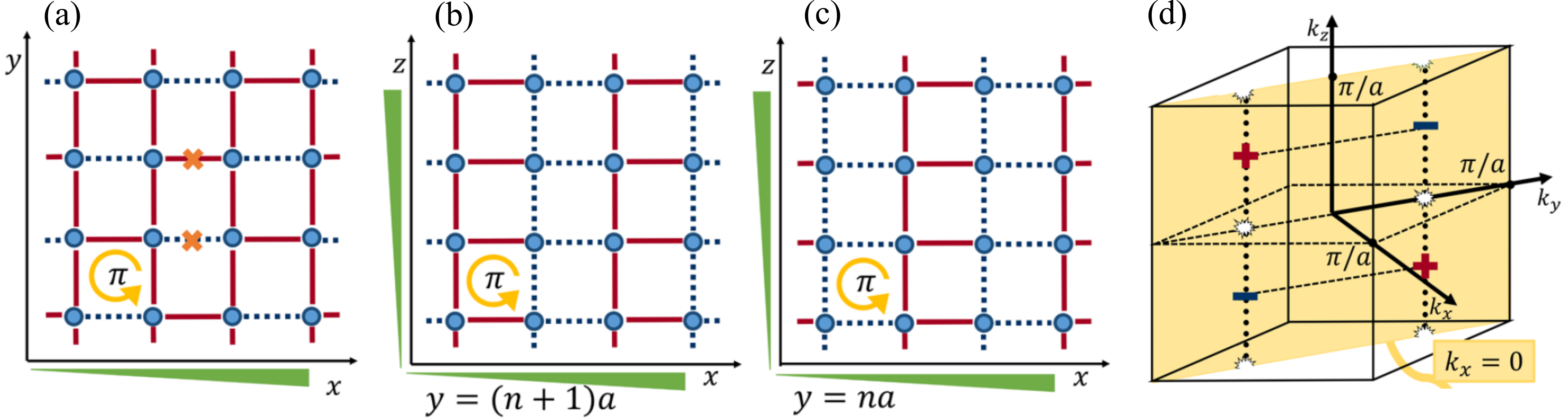}
\caption{(Color online) The hopping along the $(x,y)$ plane (a),
the $(x,z)$ plane (b-c), where
dashed and solid lines describe the hopping with an extra
phase being $\pi$ and 0, respectively. (d) First Brillouin zone
with $\pm$ denoting the Weyl points.
Adapted from Ref. [\cite{Ketterle2015PRL}].}
\label{WeylKetterle2}
\end{figure*}

The Hamiltonian can be transformed into a spatial independent form by the transformation $|j\rangle\rightarrow e^{i{\bf k}_j\cdot {\bf r}}|j\rangle$, yielding
\begin{equation}
H=\sum_j\frac{({\bf p}-{\bf k}_j)^2}{2m}|j\rangle \langle j|+H_{al},
\end{equation}
where the atom-laser coupling
\begin{eqnarray}
H_{al}=\Omega^{(1)}\sum_{j=1}^4 \left(e^{i\phi_{j,j+1}}|j+1\rangle \langle j|+H.c. \right) \nonumber \\
+\Omega^{(2)}\sum_{j=1}^2 \left(e^{i\phi_{j,j+2}}|j+2\rangle \langle j|+H.c. \right)
\end{eqnarray}
with the indices taking modulo 4.

By choosing a phase $\phi_{j,j+1}=\pi/4$ and $\phi_{j,j+2}=(j-1)\pi$, the Hamiltonian $H_{al}$ has
a symmetry $Q H_{al}Q^{-1}=H_{al}(j\rightarrow j+2)=H_{al}$. Since $Q^2=1$, one can write the Hamiltonian
in block form in the eigenstates of $Q$, that is, $\{|a_+\rangle,|b_+\rangle,|a_-\rangle,|b_-\rangle \}$ with
$|a_\pm\rangle =\frac{1}{\sqrt{2}} (|1\rangle \pm |3\rangle)$ and
$|b_\pm\rangle =\frac{1}{\sqrt{2}} (|2\rangle \pm |4\rangle)$. The block Hamiltonian is
\begin{equation}
\tilde{H}_{al}=\left(
                 \begin{array}{cc}
                   \Omega^{(2)}\sigma_z+\sqrt{2}\Omega^{(1)}\sigma_x & 0 \\
                   0 & \Omega^{(2)}\sigma_z+\sqrt{2}\Omega^{(1)}\sigma_y \\
                 \end{array}
               \right),
\end{equation}
where the upper and lower block matrices correspond to $Q=1$ and $Q=-1$, respectively.
The eigenenergy is $E_{al}=\pm\sqrt{|\Omega^{(2)}|^2+2|\Omega^{(1)}|^2}$. Since the
ground state is doubly degenerate, one can obtain an effective Hamiltonian by projecting
$\sum_{j}{\bf p}\cdot {\bf K}_j|j\rangle\langle j|$ onto the ground state subspace of $H_{al}$,
leading to the effective Hamiltonian (\ref{SpielmanHam}) with $\lambda=\kappa_\perp \cos(2\theta)/(2m)$
when $\kappa_\perp/\kappa_{||}=\Omega^{(2)}/(2\Omega^{(1)})$.

\subsubsection{Weyl semimetals with laser-assisted tunneling}
The scheme introduced in Ref.[\cite{Spielman2012PRL}] can realize a metal phase with a single Weyl point
in the spin space with finite density of states, instead of a semimetal phase with zero density of states.
In 2015, Dub\v{c}ek and coworkers proposed a distinct experimental scheme to engineer a Weyl semimetal
with ultralcold atomic gases via laser-assisted tunneling~\cite{Ketterle2015PRL}. The approach is an extension
of their experimental realization of the 2D Harper Hamiltonian for the flux per plaquette
being $1/2$~\cite{Ketterle2013PRL}, that is,
\begin{equation}
H_{\alpha=1/2}({\bf k})=-2\left[J_y\cos(k_y a)\sigma_x+K_x\sin(k_x a)\sigma_y\right].
\end{equation}
There are two Dirac points in the Hamiltonian. Here, $\sigma_\nu$ ($\nu=x,y$) are
defined in the pseudospin space consisting of two lattice sites constituting a unit
cell since the flux per plaquette is $1/2$ [see Fig.~\ref{WeylKetterle2}(a)]. When the third direction ($z$ direction) is included,
if the hopping along this direction is the same for the two sites, the Hamiltonian
reads
\begin{eqnarray}
H_{N}({\bf k})=&&-2\left[J_y\cos(k_y a)\sigma_x+K_x\sin(k_x a)\sigma_y \right. \nonumber \\
&& \left. +J_z\cos(k_z a)\sigma_0 \right].
\end{eqnarray}
Clearly, there are no Weyl points in this Hamiltonian. However, if A and B sites have the opposite
tunneling along the $z$ direction as shown in Fig.~\ref{WeylKetterle2}(b-c), the Hamiltonian becomes
\begin{eqnarray}
H_{W}({\bf k})=&&-2\big[J_y\cos(k_y a)\sigma_x+K_x\sin(k_x a)\sigma_y \nonumber \\
&& -K_z\cos(k_z a)\sigma_z\big].
\label{HLN}
\end{eqnarray}
This Hamiltonian exhibits four Weyl points at $(k_x,k_y,k_z)=(0,\pm \pi/2a,\pm \pi/2a)$
in the first Brillouin zone as shown in Fig.~\ref{WeylKetterle2}(d).
The Fermi arc appears when the open boundaries are applied along the $(x-y)$ direction.

To engineer the Hamiltonian, an experimental setup with cold atoms was introduced~\cite{Ketterle2015PRL}.
In this scheme, identical sufficiently strong
linear gradient potential along the $x$ and $z$ direction are applied so that the tunneling is prohibited
along these two directions. Meanwhile, two far-detuned Raman laser beams are used to
restore the hopping along these directions. For the resonant case (the frequency difference between
the two laser beams equals the on-site energy tilt between two nearest-neighbor sites), the
Hamiltonian is given by
\begin{eqnarray}
&&H_{3D}=-\sum_{m,n,.}(K_x e^{-i\Phi_{m,n,l}}\hat{a}_{m+1,n,l}^{\dagger}\hat{a}_{m,n,l}+  \\
&&J_y\hat{a}_{m,n+1,l}^\dagger\hat{a}_{m,n,l}
+K_ze^{-i\Phi_{m,n,l}}\hat{a}_{m,n,l+1}^\dagger\hat{a}_{m,n,l}+H.c.), \nonumber
\label{HREAL}
\end{eqnarray}
where $\hat{a}_{m,n,l}$ ($\hat{a}_{m,n,l}^\dagger$) is the annihilation (creation) operator on the
site $(m,n,l)$, and $\Phi_{m,n,l}=\delta{\bf k}\cdot{\bf R}_{m,n,l}=m\Phi_x+n\Phi_y+l\Phi_z$ are
the phases introduced by the laser-assisted tunneling. Here, $\delta{\bf k}={\bf k}_1-{\bf k}_2$ with
${\bf k}_1$ and ${\bf k}_2$ being the wave vectors of two Raman laser beams, so the phase can be
controlled by tuning the relative orientation between two laser beams. If the direction is chosen
such that
$(\Phi_x,\Phi_y,\Phi_z)=\pi(1,1,2)$ giving $\Phi_{m,n,l}=(m+n)\pi$ (modulo $2\pi$), the scheme leads to the Hamiltonian (\ref{HLN}). Specifically, the hopping
along the $x$ direction is staggered with $1$ and $-1$, respectively [see Fig.~\ref{WeylKetterle2}(a)],
leading to A-B sublattice structure. The hopping along the $z$ direction is $K_z$ and $-K_z$ for
A-B sublattices, respectively.

\subsubsection{Weyl points, structured (type-II) Weyl points and Weyl spin-orbit coupling in optical lattices}
The proposals discussed above realize only the traditional (type I) Weyl point~\cite{Spielman2012PRL,Ketterle2015PRL}.
In 2016, Xu and Duan proposed another experimental scheme for realizing both traditional (type-I) and structured (type-II) Weyl points~\cite{Yong2016PRATypeII},
based on an experimental setup used for implementing a 2D spin-orbit coupling in the hyperfine level space~\cite{Xiongjun2015Sci,Xiongjun2014PRL}.
In this scheme, the Lifshitz phase transition from
traditional (type-I) to structured (type-II) Weyl points can be readily achieved by controlling
a two-photon detuning. In addition, since the Weyl point is realized in the hyperfine level space, the scheme naturally realizes the 3D Weyl spin-orbit coupling.

\begin{figure*}[t]
\includegraphics[width=\textwidth]{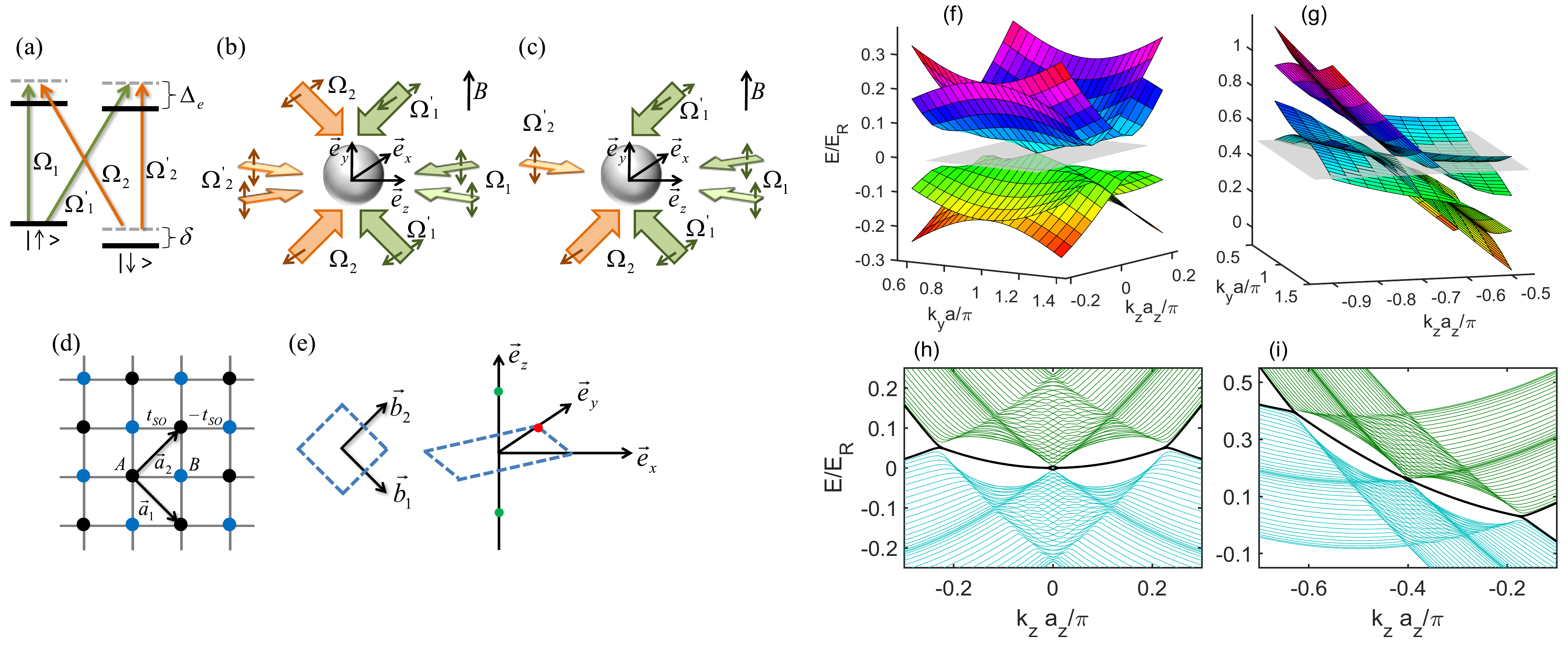}
\caption{(Color online) Sketch of laser configurations for realizing the Hamiltonian (\ref{TypeIILattice})
(a)(b) and a simpler one (a)(c). In the scheme, the magnetic field is along the $y$
direction and the frequency of the laser beams for $\Omega
_{1}$ ($\Omega _{2}$) is the same as the laser beam for $\Omega _{1}^\prime$ ($%
\Omega _{2}^\prime$). $\protect\delta$ denotes the two-photon detuning. The orientation of linear polarization of laser
beams is denoted by the double arrows.
(d) The crystal structure in the $(x,y)$ plane with A and B sites in each unit cell.
(e) The first Brillouin zone in the $(x,y)$ plane and the location of Weyl points denoted by the green and red circles.
${\bf b}_1$ and ${\bf b}_2$ denote the reciprocal unit vectors.
(f)(g) Energy spectra around
a type-I and structured (type-II) fourfold degenerate point as a function of $(k_y,k_z)$ for $%
k_xa=0$.
(h)(i) Energy spectra as a function of $k_z$ for $k_y=0$
in an open geometry along the $x$ direction, illustrating the surface states (denoted by the black lines)
connecting two Weyl points
in the cases with type-I or structured (type-II) Weyl points.
Adapted from Ref. [\cite{Yong2016PRATypeII}].}
\label{YongSingletypeII}
\end{figure*}

In this proposal, two pairs of Raman laser beams are employed to produce the Weyl point (or 3D Weyl
spin-orbit coupling).
Each pair of Raman laser beams has a pair of Rabi frequencies $[\Omega _{1}=\Omega _{10}\sin
(k_{Lx}r_{x})e^{-ik_{L{z}}r_{z}/2},\Omega _{2}=\Omega _{20}\cos (k_{L{y}%
}r_{y})e^{ik_{L{z}}r_{z}/2}]$ and $[\Omega _{1}^{\prime }=\Omega _{10}\sin
(k_{L{y}}r_{y})e^{-ik_{L{z}}r_{z}/2},\Omega _{2}^{\prime }=i\Omega _{20}\cos
(k_{L{x}}r_{x})e^{ik_{L{z}}r_{z}/2}]$, respectively [see Fig.~\ref{YongSingletypeII}(a-c) for
laser configurations]. This gives us the Hamiltonian
\begin{equation}
H^{\prime }=\frac{\mathbf{p}^{2}}{2m}+\sum_{\nu =x,y}V_{\nu }\sin
^{2}(k_{L\nu }r_{\nu })+h_{z}\sigma _{z}+V_{SO}
\label{TypeIILattice}
\end{equation}%
with the momentum operator $\mathbf{p}=-i\hbar \nabla $, the mass of atoms $m$,
the Zeeman field $h_{z}$ and the optical lattice strength $V_{\nu }$ ($\nu =x,y$)
with the lattice constant being $a_{\nu }=\pi /k_{L\nu }$ along the $\nu $
direction. Here, $V_{SO}$ characterizes the laser-induced spin-orbit coupling, that is,
\begin{equation}
V_{SO}=\Omega _{SO}(M_{x}+iM_{y})e^{ik_{Lz}r_{z}}|\uparrow \rangle
\left\langle \downarrow \right\vert +\text{H.c}.
\end{equation}%
with $M_{x}=\sin (k_{Lx}r_{x})\cos (k_{Ly}r_{y})$, $M_{y}=\sin
(k_{Ly}r_{y})\cos (k_{Lx}r_{x})$, and $\Omega _{SO}=\Omega_{10}^*\Omega_{20}/\Delta_e$, where
$\Delta_e$ is the detuning.
Applying a unitary transformation with
$U=e^{-ik_{Lz}r_{z}/2}|\uparrow\rangle\langle\uparrow|+e^{ik_{Lz}r_{z}/2}|\downarrow\rangle\langle\downarrow|$
gives us a position independent Hamiltonian $H=UH^\prime U^{-1}$, that is,
\begin{equation}
H=\frac{\hbar ^{2}k_{z}^{2}}{2m}+\tilde{h}_{z}\sigma _{z}+H_{2D},
\end{equation}%
where $k_{z}=p_{z}/\hbar $, $\tilde{h}_{z}=\hbar ^{2}k_{L{z}}k_{z}/(2m)+h_{z}
$, and the Hamiltonian $H_{2D}$ in the $(x,y)$ plane is
\begin{eqnarray}
H_{2D} &=&\sum_{\nu =x,y}\left[ \frac{p_{\nu }^{2}}{2m}+V_{\nu }\sin
^{2}(k_{L\nu }r_{\nu })\right]   \notag \\
&+&\left[ \Omega _{SO}(M_{x}+iM_{y})|\uparrow \rangle \left\langle
\downarrow \right\vert +\text{H.c.}\right] .
\label{H2D}
\end{eqnarray}
$H_{2D}$ is the Hamiltonian that characterizes a Chern insulator, which is
a simple extension of the Hamiltonian proposed in Ref.[\cite{Xiongjun2014PRL}]
and has recently been experimentally implemented~\cite{Xiongjun2018PRL}.

The tight-binding Hamiltonian is given by
\begin{eqnarray}
H_{TB}&&=\sum_{k_{z}}\sum_{{\bf x}}[ \tilde{h}_{z}\hat{c}_{k_{z},{\bf x}}^{\dagger }\sigma_z\hat{c}_{k_{z},{\bf x}}
+\sum_{\nu=x,y} (-t_{\nu}\hat{c}_{k_{z},{\bf x}}^{\dagger }\hat{c}_{k_{z},{\bf x}+{\bf g}_\nu}
\nonumber \\
&&+(-1)^{j_x+j_y}t_{SO\nu}\hat{c}_{k_{z},{\bf x}}^{\dagger }\sigma_\nu\hat{c}_{k_{z},{\bf x}+{\bf g}_\nu}+\text{H.c.}) ]
+H_z,
\label{WeylXuTB}
\end{eqnarray}
where $\hat{c}^\dagger_{k_{z},{\bf x}}=(
\begin{array}{cc} \hat{c}^\dagger_{k_{z},{\bf x},\uparrow} & \hat{c}^\dagger_{k_{z},{\bf x},\downarrow}
\end{array}
)$ with $\hat{c}_{k_{z},{\bf x},\sigma }^{\dagger }$ ($\hat{c}_{k_{z},{\bf x},\sigma }$) being the
creation (annihilation) operator and ${\bf x}=j_{x}a_x{\bf e}_x+j_{y}a_y{\bf e}_y$,
${\bf g}_\nu=a_\nu{\bf e}_\nu$,
$H_z=\sum_{k_z}\sum_{\bf x}\frac{\hbar ^{2}k_{z}^{2}}{(2m)}\hat{c}_{k_{z},{\bf x}}^{\dagger }\hat{c}_{k_{z},{\bf x}}$,
and
$t_\nu$ and $t_{SO\nu}$ denote the tunneling and spin-orbit coupling
strength along the $\nu$ direction, respectively.

To see that the Hamiltonian can host Weyl points, let us apply the transformation
$(-1)^{j_x+j_y}\hat{c}_{k_z,{\bf x},\uparrow}\rightarrow \hat{a}_{k_z,{\bf x},\uparrow}$
and $\hat{c}_{k_z,{\bf x},\downarrow}\rightarrow \hat{a}_{k_z,{\bf x},\downarrow}$, resulting in a simpler Hamiltonian,
\begin{eqnarray}
&&H_{TB}=\sum_{k_{z}}\sum_{{\bf x}}\big[ \tilde{h}_{z}\hat{a}_{k_{z},{\bf x}}^{\dagger }\sigma_z\hat{a}_{k_{z},{\bf x}}
+(\sum_{\nu=x,y} t_{\nu}\hat{a}_{k_{z},{\bf x}}^{\dagger }\sigma_z\hat{a}_{k_{z},{\bf x}+{\bf g}_\nu}
\nonumber \\
&&-it_{SOx}\hat{a}_{k_{z},{\bf x}}^{\dagger }\sigma_y\hat{a}_{k_{z},{\bf x}+{\bf g}_x}
-it_{SOy}\hat{a}_{k_{z},{\bf x}}^{\dagger }\sigma_x\hat{a}_{k_{z},{\bf x}+{\bf g}_y}+\text{H.c.}) \big] \nonumber \\
&&+H_z(\hat{c}_{k_z,{\bf x}}\rightarrow \hat{a}_{k_z,{\bf x}}),
\end{eqnarray}
where $\hat{a}^\dagger_{k_{z},{\bf x}}=(
\begin{array}{cc} \hat{a}^\dagger_{k_{z},{\bf x},\uparrow} & \hat{a}^\dagger_{k_{z},{\bf x},\downarrow}
\end{array}
)$. The Hamiltonian in momentum space takes the form
\begin{equation}
H({\bf k})=d_x\sigma_y+d_y\sigma_x+(\tilde{h}_z +h_t)\sigma_z+\frac{\hbar ^{2}k_{z}^{2}}{2m},
\end{equation}
where $h_{t}=2\sum_{\nu =x,y}t_{\nu }\cos (k_{\nu }a_{\nu })$, $%
d_{x}=2t_{SOx}\sin (k_{x}a_{x})$ and $d_{y}=-2t_{SOy}\sin (k_{y}a_{y})$. Clearly,
Weyl points appear when $d_x=0$, $d_y=0$ and $h_t+\tilde{h}_z=0$. This requires that
$(k_xa_x,k_ya_y)=(0,0),(0,\pi),(\pi,0),(\pi,\pi)$. For $t_x=t_y$, at $(k_xa_x,k_ya_y)=(0,\pi),(\pi,0)$
, the two Weyl points are both located at $%
k_{z}^{W0}a_{z}=-2m\pi h_{z}/(\hbar ^{2}k_{L{z}}^{2})$. When $(k_{x}a_{x},k_{y}a_{y})=(0,0)$ or $(\pi,\pi)$,
there appear two degenerate points at $k_{z}^{W\pm
}a_{z}=2m\pi (\pm 4\bar{t}-h_{z})/(\hbar ^{2}k_{L{z}}^{2})$ with $\bar{t}%
=(t_{x}+t_{y})/2$. Because of the presence of $\frac{\hbar ^{2}k_{z}^{2}}{2m}$, near a Weyl point,
e.g., ${\bf k}^{W0}=(0,\pi /a_{y},k_{z}^{W0})$,
\begin{equation}
H(\mathbf{q})\sim (v_{0}q_{z}+v_{z}q_{z}\sigma _{z}+v_{x}q_{x}\sigma
_{x}+v_{y}q_{y}\sigma _{y}),
\end{equation}%
where $v_x=2t_{SOx}a_x$, $v_y=-2t_{SOy}a_y$, $v_z=\hbar^2 k_{Lz}/2m$, $v_{0}=-2h_{z}/k_{L{z}}$, and $q_\nu$ ($\nu=x,y,z$) is measured with respect to ${\bf k}^{W0}$.
When $|h_{z}|>\hbar ^{2}k_{L{%
z}}^{2}/(4m)$ (i.e., $|v_{0}|>v_{z}$), the Weyl point morphs into a structured one (type-II) as
shown in Fig.~\ref{YongSingletypeII}(g). Since
$v_0$ is controlled by tuning the two-photon detuning $\delta=2h_z$, the Lifshitz transition
between the conventional (type-I) Weyl point and structured (type-II) Weyl points can be achieved.
Notice that the transformation enlarges the Brillouin zone. In the original Hamiltonian with the A-B
sublattice [see Fig.~\ref{YongSingletypeII}(d)], there are
actually two Weyl points and one fourfold degenerate point consisting of two Weyl points with
the same topological charge [see Fig.~\ref{YongSingletypeII} (e)]. It is also important to note that
the Weyl points correspond to the Weyl spin-orbit coupling, given that $\sigma_\nu$ ($\nu=x,y,z$)
act on the hyperfine levels.

Under open boundaries along the $x$ direction, there appear surface states (Fermi arc) inside the gap connecting
the fourfold degenerate point at the center to the other two Weyl points on two sides [see Fig.~\ref{YongSingletypeII} (h-i)]. This happens for both traditional (type-I)
and structured (type-II) Weyl points.

Recently, several other schemes have been proposed to realize Weyl points and 3D Weyl
spin-orbit coupling via Raman laser beams~\cite{Xiongjun2018PRA,Chuanwei2018arXiv}. Other proposed
schemes for
realizing Weyl points in non-interacting cold atoms include
stacking 2D lattice layers with checkerboard-patterned staggered fluxes~\cite{JianHua2012PRA},
stacking one-dimensional Aubry-Andre-Harper models~\cite{sarma2015PRB}, and shaking
a face-centered-cubic optical lattice~\cite{Lijun2018PRB}. There is also a proposal for
realizing a 3D Dirac semimetal with cold atoms~\cite{YongHu2018}.

\subsection{Nodal rings and experimental realization}
\begin{figure}[t]
\includegraphics[width=3.2in]{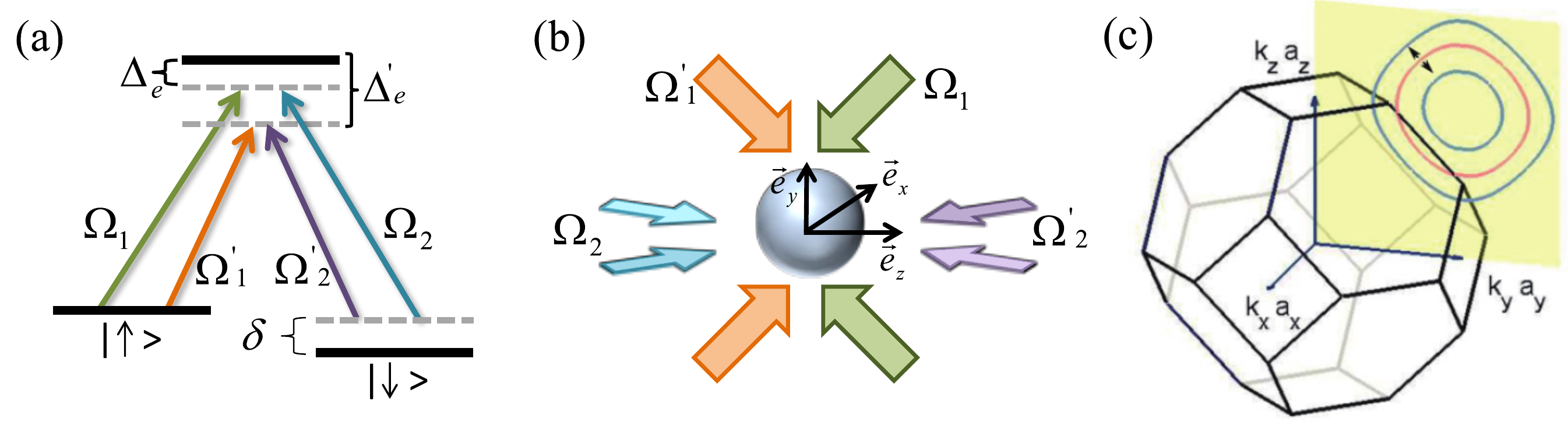}
\caption{(Color online) (a-b) Sketch of laser configurations for
realizing the Hamiltonian (\ref{NodalLineH}) with Dirac and Weyl nodal rings.
Each Raman laser beam is generated by the interference between two plane
wave lasers [see (b)]. (c) The first Brillouin zone of the system and one Dirac
(red ring) and two Weyl nodal rings (two blue rings) appearing at the $k_{x}=0$ plane.
Adapted from Ref. [\cite{Yong2016DiracRing}].}
\label{nodalring}
\end{figure}

For the nodal rings, Xu and Zhang proposed a model described by the Hamiltonian (\ref{NodalLineH})
and an experimental setup for their realization with cold atoms~\cite{Yong2016DiracRing}. The model hosts
a Dirac ring when $h_z=0$ and one or two Weyl nodal rings when $h_z \neq 0$ [see Fig.~\ref{nodalring}(c)].
To engineer such
a Hamiltonian with cold atoms, two independent sets of Raman laser
beams are proposed to couple two hyperfine states (see laser configurations in Fig.~\ref{nodalring}(a-b)).
One pair of Raman laser
beams has the Rabi frequencies $\Omega _{1}=\Omega _{10}\cos
(k_{Ry}r_{y})e^{-ik_{Rz}r_{z}/2}$ and $\Omega _{2}=i\Omega _{20}\sin
(k_{Rx}r_{x})e^{ik_{Rz}r_{z}/2}$, and the other pair has $\Omega _{1}^{\prime }=\Omega _{10}^{\prime }\cos
(k_{Ry}r_{y})e^{ik_{Rz}r_{z}/2}$ and $\Omega _{2}^{\prime }=i\Omega
_{20}^{\prime }\sin (k_{Rx}r_{x})e^{-ik_{Rz}r_{z}/2}$, respectively.
These Raman laser beams produce the spin-dependent lattice with $%
\Omega _{SO}=2\Omega $ and $a_\nu=\pi/k_{R\nu }$ with $\Omega=|\frac{\Omega_{10}\Omega_{20}}{\Delta_e}|$ in Eq.~(\ref{NodalLineH}) when $%
\Omega_{10} =\Omega ^{\prime }_{10}$ and $%
\Omega_{20} =\Omega ^{\prime }_{20}$ as well as the spin-independent lattice
along the $x$ and $y$ directions due to the stark effects. In addition,
another stronger optical lattices along the $x$ direction is required.
The Zeeman field is $h_{z}=\delta /2$, where $\delta$ is the two-photon detuning.
This shows that the Zeeman field can be readily tuned by controlling the two-photon detuning.

Recently, Song and coworkers theoretically proposed a simpler scheme and experimentally engineered a
setup for observing a nodal line with cold atoms~\cite{Song2018}. Instead of engineering the $\sigma_z$ term by
optical lattices, they impose
the off-diagonal spin-dependent lattice with an extra phase along the $z$ direction, which produces a $\sigma_z$
term~\cite{Yong2016PRATypeII}. Specifically, the proposed Hamiltonian is given by
\begin{eqnarray}
H=\frac{{\bf p}^{2}}{2m}-V_{x}\cos^{2}(k_{Lx}x)-V_{y}\cos^{2}(k_{Ly}y)+ \nonumber \\
\left[\Omega_{SO}\cos(k_{Lx}x)e^{i(k_{Lz}z+k_{Ly}y)}|\uparrow\rangle\langle\downarrow|+H.c.\right].
\label{SimplerNodalLineH}
\end{eqnarray}%
And the Hamiltonian in momentum space is
\begin{equation}
H({\bf k})=\frac{p_{z}^{2}}{2m}+\left(p_{z}\frac{\hbar k_{Lz}}{2m}+h_{z}-h_{t}\right)\sigma_{z}
-d_{x}\sigma_{y}
\end{equation}
with $h_t=2\sum_{\nu=x,y}t_\nu\cos(k_\nu a_\nu)$.
Clearly, the nodal ring appears as $d_x=0$ and $p_{z}\frac{\hbar k_{Lz}}{2m}+h_{z}-h_{t}=0$. Because of
the presence of $\frac{p_{z}^{2}}{2m}$, the nodal ring is not located at the same energy. Also, due to
this term, one can realize the structured (type-II) nodal ring by tuning the $h_z$. The Hamiltonian
is easy to realize and only requires the coupling of a stationary wave Raman laser beam with Rabi frequency being $\Omega_0\cos(k_{Lx}x)$
and a plane wave Raman laser beam with Rabi frequency being $\Omega_0^{\prime} e^{i(k_{Lz}z+k_{Ly}y)}$, respectively.
Song and coworkers experimentally engineered this model and probed the nodal ring by measuring
the average spin polarization over $k_z$ for a different
parameter~\cite{Song2018}.

\subsection{Weyl exceptional rings}

\begin{figure}[t]
\includegraphics[width=3.2in]{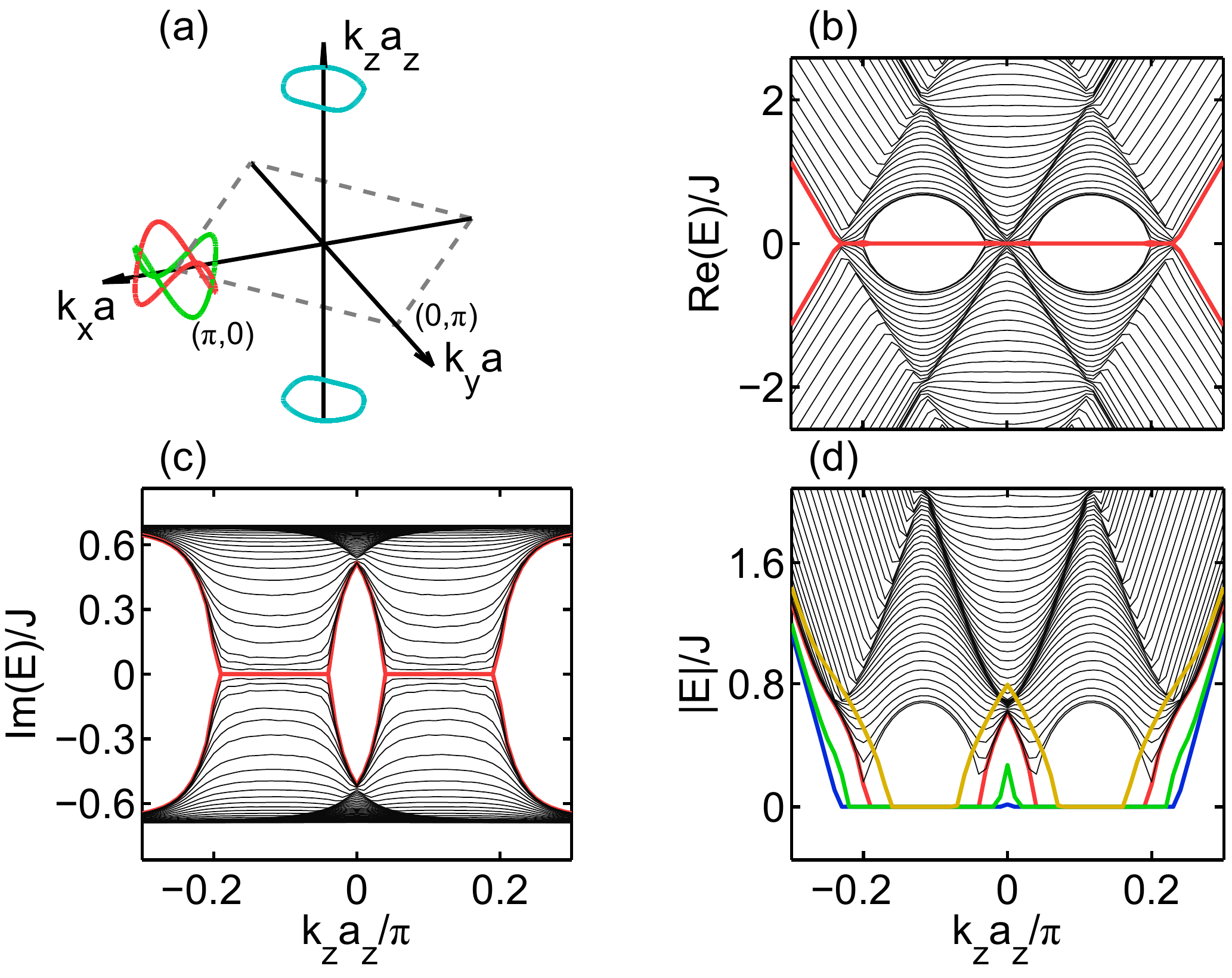}
\caption{(Color online) (a) Weyl exceptional rings in the Hamiltonian (\ref{WERLattice}).
Real part (b), imaginary part (c) and absolute values (d) of the energy spectrum
as a function of $k_z$ for $k_y=0$
with an open boundary along the $x$ direction. $\gamma=0,0.35J,0.7J,0.86J$ are plotted
as blue, green, red and yellow lines, respectively.
Reproduced from Ref. [\cite{Xu2017PRL}].}
\label{WERCold}
\end{figure}

To realize a Weyl exceptional ring in cold atom systems, one need to engineer a non-Hermitian term
representing a particle gain or loss in a system with Weyl points. For the Weyl metal described by the Hamiltonian (\ref{WeylXuTB}), such a decay term representing an atom loss $-2i\gamma$ for spin-down atoms may be produced by applying a resonant optical beam to drive atoms in the spin-down state out of a weak trap~\cite{Xu2017PRL}, which have been realized in an experiment with $^{6}$Li atom gases~\cite{LuoleArxiv}. Another considered approach is using a radio frequency pulse to excite atoms in the spin-down level to another irrelevant level $|3\rangle$ so that an effective decay for spin-down atoms is generated when atoms in $|3\rangle$ experience a loss by applying an antitrap~\cite{Xu2017PRL}.

The resulted Hamiltonian is given by
\begin{eqnarray}
H = && \sum_{k_{z},\mathbf{x}} \Big[ (\tilde{h}_{z}+i\gamma)\hat{c}_{k_{z},%
\mathbf{x}}^{\dagger }\sigma_z\hat{c}_{k_{z},\mathbf{x}} + \sum_{\nu=x,y} (-J%
\hat{c}_{k_{z},\mathbf{x}}^{\dagger }\hat{c}_{k_{z},\mathbf{x}+ a\mathbf{e}%
_\nu}  \notag \\
&& +(-1)^{j_x+j_y}J_{SO\nu}\hat{c}_{k_{z},\mathbf{x}}^{\dagger }\sigma_\nu%
\hat{c}_{k_{z},\mathbf{x}+ a\mathbf{e}_\nu }+H.c. )+h_0 \Big],
\label{WERLattice}
\end{eqnarray}
where $\gamma$ denotes the decay strength, and $h_0=[-i\gamma+\hbar ^{2}k_{z}^{2}/(2m)]\hat{c}%
_{k_{z},\mathbf{x}}^{\dagger }\hat{c}_{k_{z},\mathbf{x}}$. For other terms,
see the description for the Hamiltonian (\ref{WeylXuTB}). In momentum space, the Hamiltonian reads
\begin{equation}
H(\mathbf{k})=(\tilde{h}_{z}+i\gamma
)\sigma_{z}-h_{t}\tau_{x}+\tau_{y}(-b_{x}\sigma_{x}+b_{y}\sigma_{y}),
\label{HTBk}
\end{equation}
where without loss of generality the spin-independent term $h_0$ is neglected.
It gives the eigenvalues $E_{\theta_\pm}(\mathbf{k})=\sqrt{b_{\pm}^2-\gamma^2+2ib_{z%
\pm}\gamma}=\sqrt{A_{\pm}(\mathbf{k})}e^{i\theta_{\pm}/2}$, where $A_{\pm}(%
\mathbf{k})=\sqrt{(b_{\pm}^2-\gamma^{2})^{2}+4b_{z\pm}^{2}\gamma^{2}}$ with $%
b_{\pm}^2=b_{x}^{2}+b_{y}^{2}+b_{z\pm}^{2}$ and $b_{z\pm}=\pm h_{t}+\tilde{h}%
_{z}$, and $\theta_\pm$ are defined by $%
\cos\theta_{\pm}=(b_{\pm}^2-\gamma^{2})/A_{\pm}(\mathbf{k})$ and $%
\sin\theta_{\pm}=2b_{z\pm}\gamma/A_{\pm}(\mathbf{k})$.
The four Weyl exceptional rings appear when $b_{z\pm}=b_{\pm}^2-\gamma^{2}=0$,
as shown in Fig.~\ref{WERCold} (a).

Under open boundary conditions along the $x$ direction, zero energy surface states
are observed even with the non-Hermitian term [see Fig.~\ref{WERCold}(b-d)]. While the energy becomes
complex for the non-Hermitian system, for $k_y=0$, only the states with
zero absolute energy are associated with the surface states, which connect the Weyl
exceptional ring at the center with the rings on the left and right sides ($h_0$ has been
neglected for clarity). These states
are doubly degenerate and are located on the left and right surface,
respectively. As the decay increases, the range of the surface states along $k_z$ shrinks
due to the structure developed along the $k_z$ direction.

\section{Other topological gapless matters in cold atoms}
Gapless points not only can occur in the two level space, but also appear in the space with more than two levels.
Recently, 3D high spin fermions with higher Chern numbers have been predicted~\cite{Kane2016PRL,Bernevig2016Sci}.
Such fermions are described by $H={\bf k}\cdot {\bm S}$ with $\bm S$ being angular
momentum matrices corresponding to high spins. For triply degenerate points, Zhu and coworkers~\cite{ShiLiang2017PRA}
constructed a 3D tight-binding model
and proposed an experimental scheme to realize the triply degenerate points in cold atom systems. Hu and coworkers~\cite{Haiping2018PRL}
predicted a new triply degenerate point with different topological charges induced by the
spin-tensor-momentum coupling and reported an experimental scheme for their realization in
cold atom optical lattices.
Fulga and coworkers~\cite{Burrello2018PRB} reported a proposal for realizing topologically protected crossings of three energy bands on a 3D Lieb lattice.
Further, Mai and coworkers~\cite{ShiLiang2018} proposed a double triple point
emerging in topological metal bands as a spin-1 generalization of double-Weyl points.

Moreover, cold atoms can be used to mimic the gapless phenomena in higher than three
dimensions, such as Weyl surfaces and Yang monopoles~\cite{Lian2016PRB,Lian2017PRB} in five dimensions and
4D nodal rings~\cite{Yanbin2018PRB} in four dimensions. In particular, recently, Sugawa and coworkers~\cite{Spielman2018Science} reported their measurement of the second Chern number of the Yang monopole using the
hyperfine levels of atomic Bose-Einstein condensates.

\section{Summary and perspectives}
Over the past decade, cold atoms have witnessed a rapid development in simulating topological
matters. A number of topological phenomena that have been long-sought after in solid-state
materials and particle physics have been experimentally realized in cold atoms, such as the Haldane model~\cite{Jotzu2014Nature},
the Thouless pump~\cite{Nakajima2016,Bloch2016}, the 4D quantum Hall effect~\cite{Goldman2015PRL,Bloch2018Nature} and the
Yang monopole~\cite{Spielman2018Science}. New topological gapless phenomena
have also been predicted in cold atom systems, such as structured (type-II) Weyl points~\cite{Xu2015PRL} and Weyl exceptional rings~\cite{Xu2017PRL}.
To date, there have been a number of experimental proposals for realizing Weyl points,
structured (type-II) Weyl points, Dirac points, nodal rings, structured (type-II) nodal
rings, Weyl exceptional rings, triple points and other 3D gapless phenomena with cold atoms.
In particular, a recent experiment has
made a significant breakthrough for engineering and observing the nodal line in cold atoms~\cite{Spielman2018Science}.

In addition, there has been remarkable progress in engineering both 1D and 2D spin-orbit coupling
in Fermi gases, which paves the way for observing these 3D gapless phenomena in qusiparticle spectra of spin-orbit coupled Fermi
superfluids. However, it still remains a significant challenge for achieving the spin-orbit coupled
Fermi superfluids due to the heating issue in alkali atoms while producing the spin-orbit coupling,
which makes the sufficient low temperature required for superfluidity hard to reach~\cite{HuiZhai2015}.
Remarkably, Ye and coworkers recently reported
the experimental realization of the spin-orbit coupling in an optical clock using a direct optical
clock transition between two electric orbital states
in $^{87}$Sr atoms, instead of using the Raman coupling which leads to heating. Since this platform is
not plagued by the heating issue, it may be promising for achieving spin-orbit coupled superfluids.

For non-interacting systems, the challenge does not generally lie in heating.
Another more significant challenge (for both non-interacting systems and superfluids) arises. That is how to
measure the Weyl points and nodal rings. Generically, one can measure the
spectrum using momentum-resolved radio-frequency spectroscopy, similar to the ARPES in solid
states. However, in 3D, such a measurement involves stacking the information of
different layers while making time-of-flight measurement; such stacking would smear out the
single layer information and raise a significant challenge for reconstructing the spectrum.
Certainly, for a specific system, one may design some special technique to reconstruct the
information, for example, in the experimental observation of a nodal line, the average spin
polarization over $k_z$ for a different
parameter can reflect the polarization for different $k_z$. Another possible choice is to measure the
Fermi arc for Weyl semimetals and drumhead surface states for Weyl nodal semimetals by
loading atoms into a box potential. For an ideal semimetal, since only Fermi arc or flat surface
states contribute nonzero density of states at zero energy, this allows us to obtain the
spectrum of surface states without the bulk effects. For a Weyl semimetal, one may also consider measuring
the topological anomalous Hall effect~\cite{YongHu2018}.

\begin{acknowledgments}
We appreciate the collaboration with C. Zhang, L.-M. Duan, F. Zhang, S.-T. Wang, R.-L. Chu,
Y. Hu and Y.-B. Yang.
We thank Q.-B. Zeng, Y.-B. Yang and Y.-L. Tao for helpful discussions and critical reading
of the manuscript.
This work was supported by the start-up program of Tsinghua University and
the National Thousand-Young-Talents Program.
\end{acknowledgments}

\end{document}